\documentclass[prd,showpacs,superscriptaddress,notitlepage,longbibliography]{revtex4-1}
\usepackage[usenames,dvipsnames]{color}
\usepackage[T1]{fontenc}
\usepackage{amsmath}
\usepackage{amssymb}
\usepackage{latexsym}
\usepackage{enumerate}
\usepackage{bbm}
\usepackage{amsthm}
\usepackage[colorlinks]{hyperref}
\usepackage{graphicx}
\usepackage{dcolumn}
\usepackage{bm}
\hypersetup{citecolor=red, linkcolor=blue}
\usepackage{subfigure}
\usepackage{footnote}
\usepackage{ulem}
\usepackage{mathrsfs}
\usepackage{url}
\DeclareSymbolFontAlphabet{\mathrsfs}{rsfs}
\newcommand{\scri}{\mathrsfs{I}}

\newcommand{\dd}{\mathrm{d}} 
\renewcommand{\emph}[1]{\textit{#1}}

\begin{document}

\title{Global simulations of Minkowski space-time including space-like infinity}

\author{Georgios Doulis}
\email{gdoulis@phys.uoa.gr}
\affiliation{Max Planck Institute for Gravitational Physics (Albert-Einstein-Institute).}

\author{J\"org Frauendiener}
\email{joergf@maths.otago.ac.nz}
\affiliation{Department of Mathematics and Statistics, University of Otago,
  Dunedin 9010, New Zealand.}
\affiliation{Department of Mathematics, University of Oslo, Norway.}


\vspace{1cm}

\begin{abstract}

In this work, we study linearised gravitational fields on the entire Minkowski space-time including space-like 
infinity. The generalised conformal field equations linearised about a Minkowski background are utilised 
for this purpose. In principle, this conformal representation of Einstein's equations can be used to carry 
out global simulations of Minkowski space-time. We investigate thoroughly this possibility.
\end{abstract}

\maketitle

\numberwithin{equation}{section}


\section{Introduction}
\label{sec:intro}

The recent direct observation of gravitational waves \cite{GW150914} demonstrated conclusively the predictive 
power of numerical relativity in problems of high complexity where the known analytical methods can not 
be applied. Since there are no analytical expressions for the waveforms expected from binary black hole 
mergers, like the one observed by LIGO, they were obtained numerically and then compared to the actual 
findings of \cite{GW150914}. The fact that the observed waveforms agree to a high level of accuracy with 
the theoretically expected ones marks another success of General Relativity. 

Numerical relativity experienced a rapid advance in the last couple of decades, fueled mainly by the urge 
to model binary systems of massive compact objects. The standard way of describing a system of this kind 
is as an isolated system, i.e. a compact self-gravitating astrophysical object embedded in an asymptotically 
flat space-time. Although this approximation provides a rigid basis for the analytical and numerical 
study of self-gravitating systems, it puts a huge burden especially on the numerical side since now the 
infinite span of the asymptotically flat space-time must be somehow simulated with finite computational 
resources. Trying to deal with this question several approaches have been developed over the years. 

The most common approach is based on the Cauchy formulation of Einstein's equations. In this approach the 
space-like hyper-surfaces of constant time are truncated at a finite distance where an artificial time-like 
boundary is introduced. To ensure that the resulting initial boundary value problem is mathematically 
well-posed and numerically stable appropriate boundary conditions must be imposed at this boundary. The 
boundary conditions must satisfy the constraints on every timeslice and must be absorbing, i.e. purely 
outgoing, in order to minimise the amount of spurious reflections on the boundary. During the last decade, 
following the seminal works \cite{Pretorius2005,Campanelli2006,Baker2006}, more and more sophisticated 
codes were developed based on this approach. An indication of the high quality of these numerical schemes 
is the fact that the numerically computed waveforms with which the actual waveforms detected by LIGO 
\cite{GW150914} have been compared with were modeled upon \cite{Campanelli2006}. Despite its successes, 
and the fact that there is definitely space for further improvement, the Cauchy approach has certain 
limitations emanating mainly from the fact that in General Relativity local expressions for the gravitational 
energy density and flux do not exist. This in turn makes impossible the construction of completely absorbing 
boundary conditions for boundaries that stand at finite distance. In addition, as shown in \cite{Zenginoglu2008a}, 
when gravity is coupled to (e.g. scalar or Yang-Mills) fields the location where information is extracted 
from is of essential physical importance as the decay rates of the fields depend on the location of the 
observer. One has also to bear in mind that while the Cauchy approach is very well adapted to the study 
of binary and other isolated systems, it cannot be used to answer questions that require the inclusion 
of space-time infinity into the computational domain like the feasibility of global simulations of entire 
space-times, the stability of black hole space-times, and the cosmic censorship conjecture. Strictly 
speaking, even the aim of computing the exact gravitational wave forms emitted from an isolated system 
necessitates the inclusion of the entire space-time. However, for the accuracies needed for the current 
detectors this does not seem to be an issue.

A possible way to extend the standard Cauchy approach all the way out to infinity is by combining it with 
the so-called characteristic approach \cite{Bondi1962,Sachs1962}. This approach uses light-like, instead 
of space-like, hyper-surfaces that reach all the way out to null infinity through a compactification of 
the spatial coordinate. In the resulting Cauchy-characteristic matching method \cite{Bishop1993}, the 
interior region is treated with the standard Cauchy techniques, while the distant asymptotically flat 
region is left to the characteristic approach. The two approaches are matched along a transparent, continuous 
(and at least once differentiable) time-like boundary that has the dual role of providing an outer (inner) 
boundary condition for the Cauchy (characteristic) evolution. The data flow at the boundary is two-way with 
the Cauchy and characteristic codes providing exact boundary values for each other. The construction of such 
a two-way boundary is not so trivial as the whole (analytical and numerical) setup changes at the interface. 
This complexity calls for the development of highly sophisticated techniques that could enable the construction 
of such a boundary. Although, considerable progress has been made in that direction \cite{Szilagyi&Winicour2003}, 
the current implementation of the boundary is still one-way \cite{Reisswig2009}. Specifically, absorbing 
boundary conditions are imposed on the interface between the two codes with data flowing only from the 
Cauchy to the characteristic region, but not vise versa. Thus, in its current status the Cauchy-characteristic 
matching suffers of the same limitations as the standard Cauchy method.

It seems that the inclusion of infinity into the computational domain demands a more sophisticated approach 
than just the introduction of a boundary at finite distance. Penrose's concept of conformal infinity \cite{Penrose1963} 
provides an unexpectedly simple and mathematically sound way to deal with the difficulties related to the 
infinite span of asymptotically flat space-times. In this picture the space-time metric $\bar{g}$ is conformally 
transformed $\bar{g} = \Omega^{-2} g$ with an appropriately chosen conformal factor $\Omega$. In this way 
space-time infinity is brought to a finite distance. In the conformal space-time $g$, the outer boundary 
of the computational domain (that lies now at infinity) emerges naturally as the locus where $\Omega = 0$ 
and thus it does not have to be artificially introduced. In addition, as it lies at infinity, it is completely 
absorbing and thus no boundary conditions have to be imposed there. Research in this setting has taken two 
different but closely related directions.

In the first approach the space-time is foliated with space-like hyperboloidal hyper-surfaces that reach 
null infinity in order to avoid the singular nature of space-like infinity. A number of different formulations 
of Einstein's equations in the context of the hyperboloidal foliation have appeared through the years. 
In \cite{Friedrich1981} Friedrich managed to express Einstein's equations for the conformal metric as a 
manifestly symmetric hyperbolic system that is regular all the way to null infinity. With this set of 
equations hyperboloidal initial data have been successfully evolved along null infinity, and even up to 
time-like infinity, in several different scenarios \cite{Huebner2001,Frauendiener2000,Frauendiener2002}. 
An alternative approach by Moncrief and Rinne \cite{Moncrief&Rinne2009} employs the standard ADM formulation 
to express the conformally transformed Einstein equations on hyperboloidal hyper-surfaces of constant 
mean curvature. Although formally singular, the resulting equations are actually regular at null infinity 
provided that certain regularity conditions hold there. Based on this formulation long-term stable numerical 
evolution was achieved in axial and spherical symmetry with \cite{Rinne&Moncrief2013,Rinne2014b} and without 
\cite{Rinne2010} matter. Recently, the first extensive numerical implementation \cite{VinualesPhD} of yet 
another formulation of Einstein's equations on constant mean curvature slices by Zengino\u{g}lu \cite{Zenginoglu2008b} 
has appeared with very encouraging results.

The main motivation for using hyperboloidal hyper-surfaces is the avoidance of space-like infinity. Thus, 
any approach based on a hyperboloidal foliation is only capable of evolving data along future null infinity 
and not along past null infinity as this would require to go through the singular space-like infinity. 
So, global simulations are not possible in this approach. Another implication of excluding space-like 
infinity is that phenomena, like e.g. scattering of gravitational waves, related to the inflow of gravitational 
radiation from null infinity cannot be studied. To address such kind of questions space-like infinity 
must be brought into the picture. Recently, it was proposed in \cite{Doulis&Rinne2016} that this could 
be done by extending the hyperboloidal approaches \cite{Moncrief&Rinne2009} in a way that space-like 
infinity is taken into account. The basic idea is to use a standard Cauchy evolution scheme to obtain 
data on a first hyperboloidal slice by evolving initial data that extend to space-like infinity and then 
use these data as initial data for an already existing hyperboloidal code \cite{Rinne2010}. The feasibility 
of this idea crucially depends on the possibility of constructing Cauchy initial data that are static in 
the neighbourhood of space-like infinity and non-static in the interior. The static character of the data 
close to space-like infinity guarantees that during the short Cauchy evolution the boundary condition 
imposed on the outer boundary, which is placed well inside the static region, is exact. This kind of initial 
data have been recently constructed in \cite{Doulis&Rinne2015,Doulis&Rinne2016}.

Another way of introducing space-like infinity into the conformal picture is by employing the so-called 
generalised conformal field equations presented by Friedrich in \cite{Friedrich1998}. In this approach 
the space-time is foliated with generic space-like hyper-surfaces and Einstein's equations are reformulated 
as a symmetric hyperbolic system that is regular at space-like and null infinity. The basic ingredient 
of this approach is the blowing-up of space-like infinity to a cylinder $I = [-1, 1] \times \mathbb{S}^2$ 
of finite length that serves as a link between past and future null infinity. An extremely pleasant feature 
of the generalised conformal field equations is that they reduce to an intrinsic set of evolution equations 
on $I$, i.e. $I$ is a total characteristic of the system, and thus no boundary conditions are needed there. 
In addition, although high in number, the generalised conformal field equations have an extremely simple 
form as the majority of them are ordinary differential equations and the rest can be written in a symmetric 
hyperbolic form. The first attempt to implement this approach numerically appeared in \cite{ZenginogluPhD} 
for massless axisymmetric space-times. The Otago relativity group followed with an extensive study of the 
behaviour of linearised gravitational fields on a Minkowski background \cite{Beyer2012,Doulis2013,Beyer2014a,Beyer2014b,DoulisPhD}. 
Therein, a stable and convergent code has been developed that can evolve successfully several different 
types of initial data along (and in the neighbourhood of) the cylinder $I$. Recently, the generalised
conformal field equations have been used to compute a simulation of the response of a Schwarzschild black 
hole to the impact of gravitational waves \cite{StevensPhD} on a computational domain that includes the 
interior of the horizon and parts of null-infinity, but not $I$.

In the present work the generalised conformal field equations \cite{Friedrich1998} will be used to study 
gravitational perturbations on Minkowski space-time not only in the neighbourhood of space-like infinity 
but on the entire Minkowski space-time $\mathbb{M}$. To do so, we first embed conformally $\mathbb{M}$ 
into the Einstein static universe $\mathbb{E}$, see Sec.~\ref{sec:compactification}. Then, in Sec.~\ref{sec:blow_up}, 
by rescaling appropriately this finite representation of $\mathbb{M}$ space-like infinity is blown up to 
a cylinder in accordance with Friedrich's construction. But now, the range of the equations covers the 
entire Minkowski space-time and not only the regions close to $I$. Since we explicitly make use of the 
spherical symmetry of the underlying Minkowski space-time by using polar coordinates, we have to pay the 
price that some terms of the generalised conformal field equations are now singular at the origin, see 
Sec.~\ref{sec:GCFE}. We express them as a system of first, Sec.~\ref{sec:spin2}, and second, Sec.~\ref{sec:spin2_wave}, 
order partial differential equations and choose to implement numerically the latter as it behaves better 
numerically \cite{Kreiss&Ortiz2002}. In order to provide initial data for the latter and to guarantee 
that the solutions they provide are the same, a correspondence is established between the two systems in 
Sec.~\ref{sec:first_to_second}. In Sec.~\ref{sec:num_origin}, we describe the way in which the singular 
terms of the system \eqref{spin2_wave_equation_final} have been implemented at the origin. The above setting, 
in principle, could be used to carry out global simulations of $\mathbb{M}$ if the expected degeneracy 
of the evolution equations at the interface of $I$ with null infinity $\scri$ could be somehow circumvented. 
The conformal compactification of Fig.~\ref{subfig:horizontal_compact} resulting from the rescaling \eqref{f_horizontal} 
could serve this purpose. This possibility is thoroughly investigated in Sec.~\ref{sec:num_results_hor}.


\section{Minkowski space-time}

In this section, we discuss the details of the setting in which Minkowski space-time will be used in the 
following. The finite representation of Minkowski space-time as part of the Einstein static universe will 
be briefly presented and the blowing up of space-like infinity to a finite cylinder pioneered in \cite{Friedrich1998} 
will be described.   

\subsection{Conformal compactification}
\label{sec:compactification}

In the present work, we want to perform a similar conformal transformation and coordinate change as we 
did in \cite{Beyer2012,Doulis2013}. However, this time we are not only interested in the neighbourhood 
of space-like infinity $i^0$ but in the entire Minkowski space $\mathbb{M}$. Now, we do not perform a 
coordinate inversion to get $i^0$ to be the new origin, but we use the well known conformal embedding 
of $\mathbb{M}$ into the Einstein static universe $\mathbb{E}$ to obtain a finite representation of 
$\mathbb{M}$. 

Recall that $\mathbb{E}$ is the manifold $\mathbb{R} \times \mathbb{S}^3$ with metric
\begin{equation}
 \label{einstein_metric}
  g_E = \dd T^2 - \dd R^2 - \sin^2 \!R \left(\dd \theta^2 + \sin^2 \!\theta \,\dd \phi^2 \right),
\end{equation}
where the coordinates range in the intervals 
\begin{equation}
 \label{einstein_range}
  -\infty < T < \infty, \quad 0 < R < \pi, \quad 0 < \theta < \pi, \quad -\pi < \phi < \pi.
\end{equation}

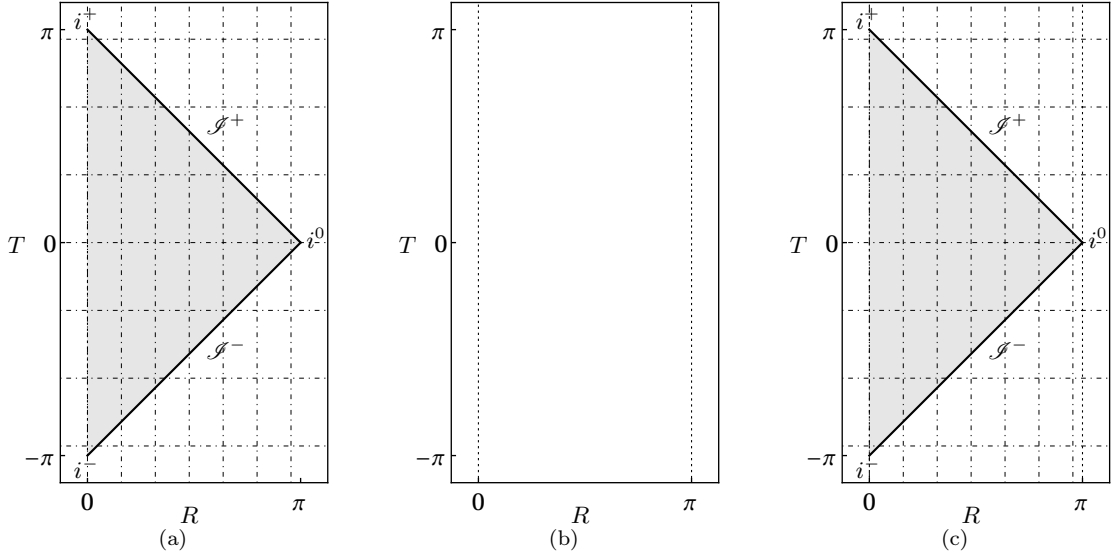
\begin{figure}[htb]
 \centering
  \subfigure[]{
   \includegraphics[scale = 0.45]{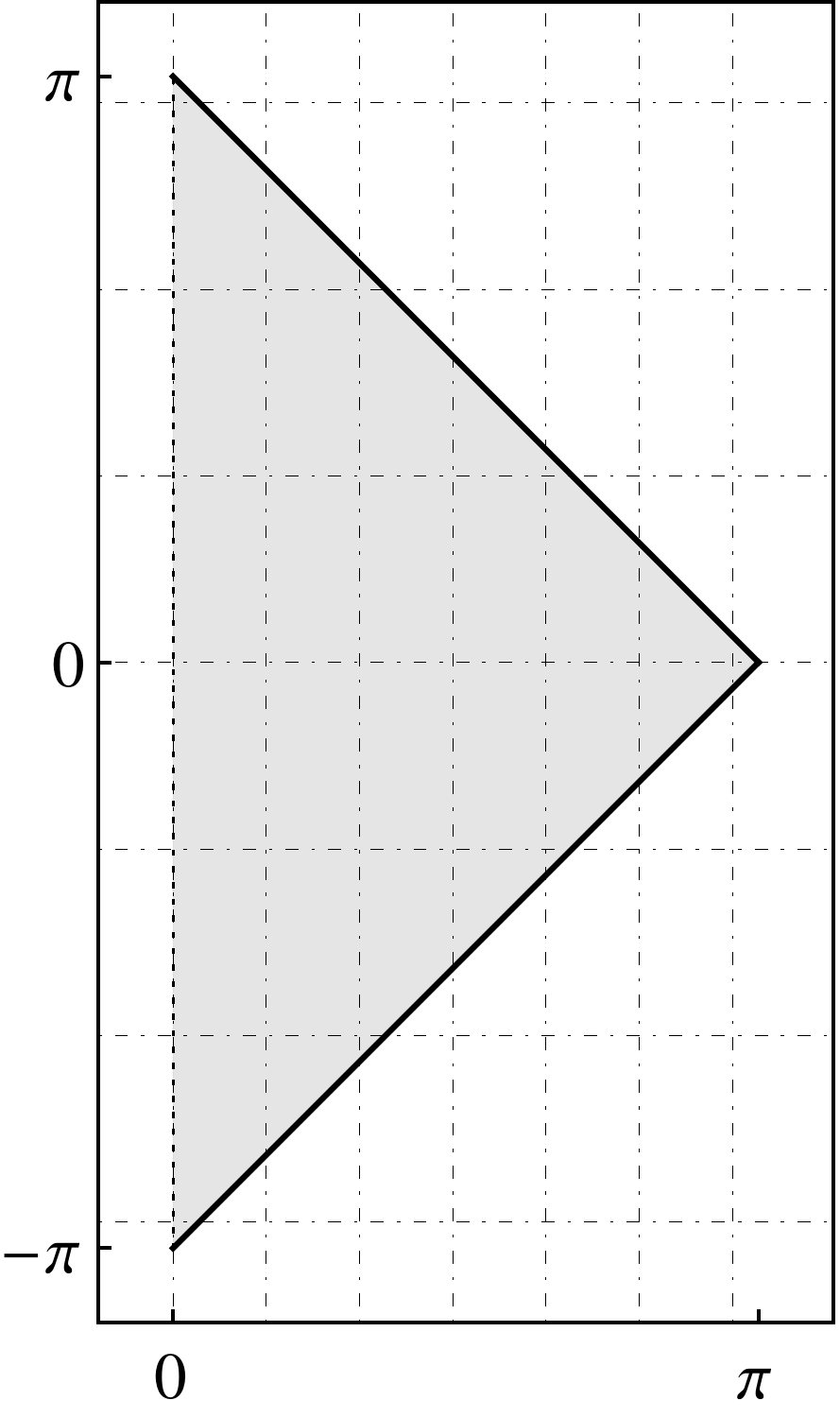}
   \label{subfig:mink_compact}
   \put(-100,15){$i^-$}
   \put(-50,60){$\scri^-$}
   \put(-12,102){$i^0$}
   \put(-50,145){$\scri^+$}
   \put(-100,185){$i^+$}
   \put(-125,100){$T$}
   \put(-60,-2){$R$}
  }
  \hspace{0.7cm}
  \subfigure[]{
   \includegraphics[scale = 0.45]{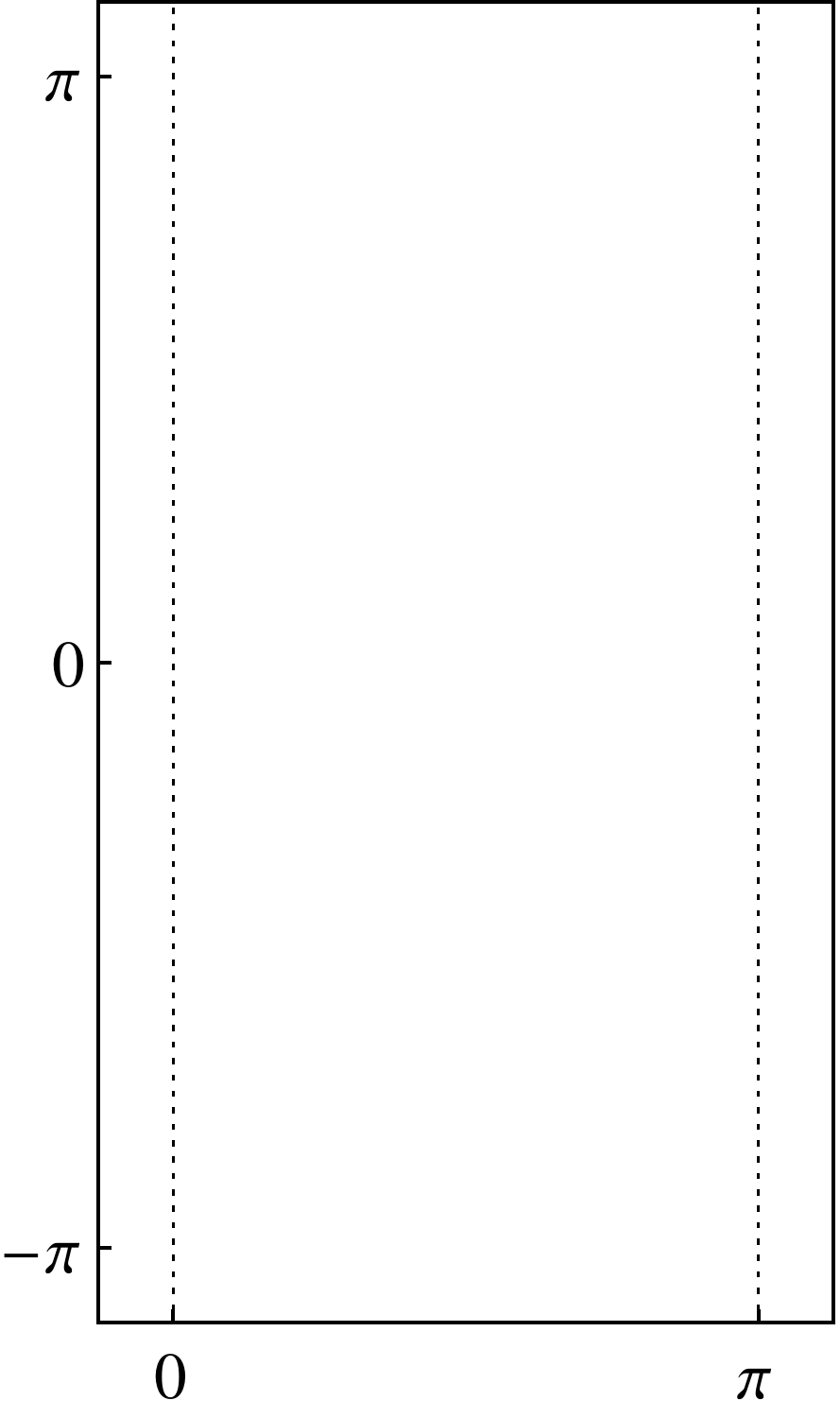}  
   \label{subfig:einstein_static}
   \put(-125,100){$T$}
   \put(-60,-2){$R$}
  }
  \hspace{0.7cm}
  \subfigure[]{
   \includegraphics[scale = 0.45]{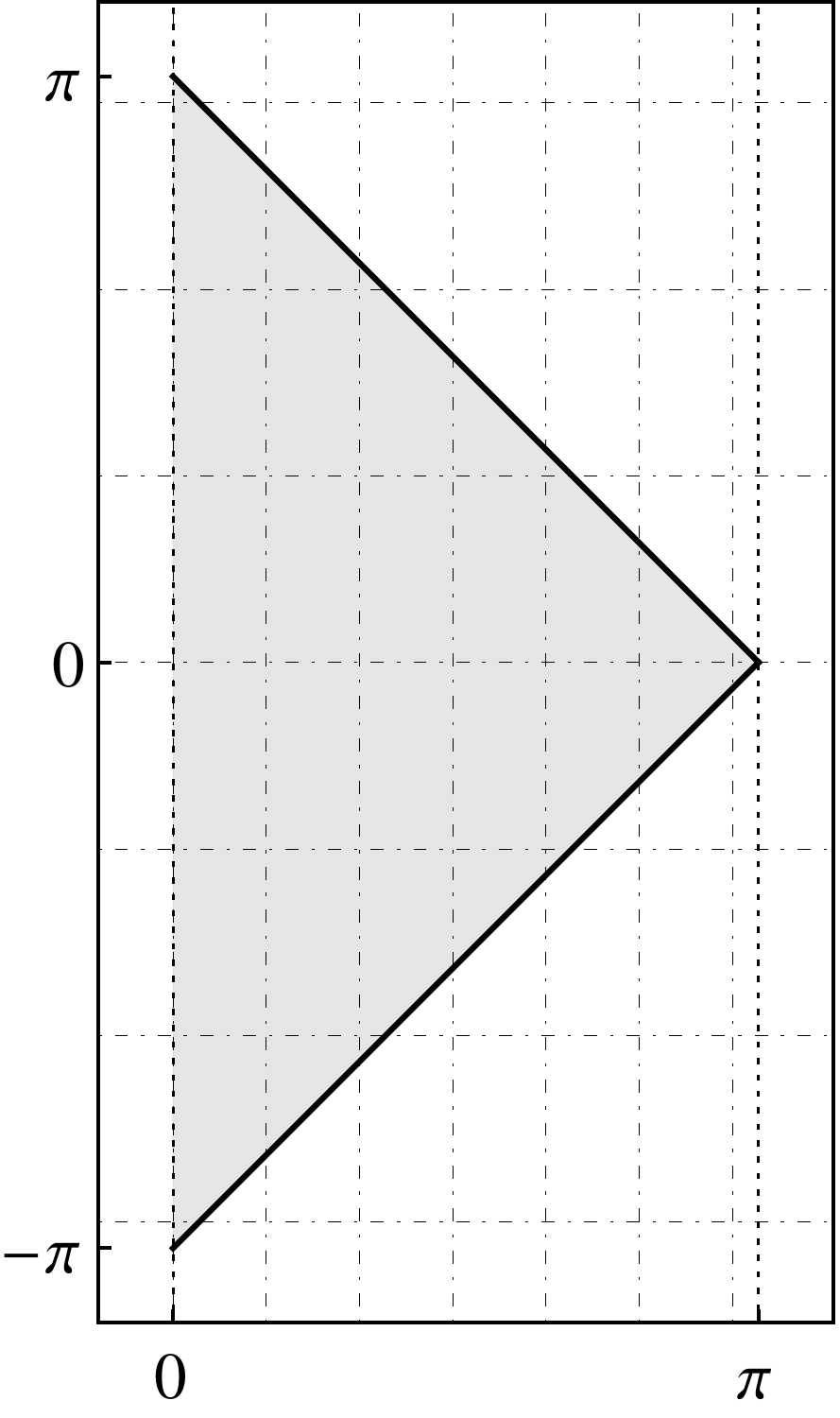}
   \label{subfig:mink_in_ein}
   \put(-100,15){$i^-$}
   \put(-50,60){$\scri^-$}
   \put(-12,102){$i^0$}
   \put(-50,145){$\scri^+$}
   \put(-100,185){$i^+$}
   \put(-125,100){$T$}
   \put(-60,-2){$R$}
  }
 \caption{Conformal compactification of Minkowski space-time into the Einstein static universe. Conformal 
 diagrams of (a) Minkowski, (b) the Einstein static universe \eqref{einstein_metric} and \eqref{einstein_range}, 
 and (c) Minkowski embedded in the Einstein static universe \eqref{mink_comp_metric} and \eqref{mink_comp_range} 
 on the $(T, R)$ plane. The usual rules apply: every point in these diagrams represents  a 2-sphere, except 
 for those on dotted lines which represent points. Points on solid lines are points at infinity. The structure 
 of conformal infinity is clearly visible.}
 \label{fig:compactification}
\end{figure}

Following \cite{Hawk&Ellis1973}, we briefly show how the entire Minkowski space-time can be conformally 
compactified and represented as a finite part of the Einstein static universe. Our starting point will 
be the Minkowski metric expressed in spherical coordinates 
\begin{equation}
 \label{mink_metric}
  \bar{g}_M = \dd t^2 - \dd r^2 - r^2 \left(\dd \theta^2 + \sin^2 \!\theta \,\dd \phi^2 \right),
\end{equation}
which when expressed in advanced $u = t + r$ and retarded $w = t -r$ null coordinates becomes
\begin{equation*}
 \bar{g}_M = \dd u\, \dd w - \frac{1}{4}(u - w)^2 \left(\dd \theta^2 + \sin^2 \!\theta \,\dd \phi^2 \right),
\end{equation*}
where $-\infty < w \leq u < \infty$. To employ Penrose's technique \cite{Penrose1963} of bringing infinity 
into finite distance, new null coordinates that assign finite values to the infinities of $u, w$ must be 
defined. A possible choice is $p = \arctan u$ and $q = \arctan w$ with $-\pi/2 < q \leq p < \pi/2$. Notice 
that the infinities $\pm \infty$ of $u, w$ have been mapped to the finite values $\pm \pi/2$ of $p, q$. 
In these coordinates the Minkowski metric takes the form
\begin{equation*}
 \bar{g}_M = \Omega^{-2} \left[4\, \dd p\, \dd q - \sin^2(p - q) \left(\dd \theta^2 + \sin^2 \!\theta \,\dd \phi^2 \right) \right],
\end{equation*}
where the conformal factor $\Omega(p,q) = 2 \cos p\, \cos q$ is positive definite in the domain $p,q \in (-\pi/2, \pi/2)$ 
and has the appropriate behaviour at infinity, i.e. $\Omega(\pm \pi/2,\pm \pi/2) = 0$. Thus, the metric 
$\bar{g}_M$ has been conformally transformed to the metric 
\begin{equation*}
 g_M = 4\, \dd p\, \dd q - \sin^2(p - q) \left(\dd \theta^2 + \sin^2 \!\theta \,\dd \phi^2 \right).
\end{equation*}
Finally, effecting the transformation $(T, R) = (p + q, p - q)$, the conformal metric $g_M$ can be brought 
into the form \eqref{einstein_metric}, namely
\begin{equation}
 \label{mink_comp_metric}
  g_M = \dd T^2 - \dd R^2 - \sin^2 \!R \left(\dd \theta^2 + \sin^2 \!\theta \,\dd \phi^2 \right),
\end{equation}
where the coordinates now must satisfy the relations 
\begin{equation}
 \label{mink_comp_range}
  -\pi < T + R < \pi, \quad -\pi < T - R < \pi, \quad -\pi < T < \pi, \quad 0 < R < \pi, \quad 0 <\theta < \pi, \quad -\pi < \phi < \pi.
\end{equation}
Conditions \eqref{mink_comp_range} restrict the whole of Minkowski space-time $\mathbb{M}$ to a finite 
open subset of the Einstein static universe $\mathbb{E}$---compare with \eqref{einstein_metric} and \eqref{einstein_range}. 
In fact, the conformal infinity of Minkowski space-time lies at the boundary of this region. The location 
of this boundary is given by the vanishing of the conformal factor $\Omega$, which in the coordinates 
$(T, R, \theta, \phi)$ reads 
\begin{equation}
 \label{Omega}
  \Omega(T, R) = 2 \cos \left(\frac{T + R}{2}\right) \cos \left(\frac{T - R}{2}\right)
\end{equation}
Thus, the structure of conformal infinity is as follows, see Fig.~\ref{fig:compactification}. Past time-like 
infinity $i^-$, space-like infinity $i^0$, and future time-like infinity $i^+$ are located at the coordinate 
points $(T, R) = (-\pi, 0)$, $(T, R) = (0, \pi)$, and $(T, R) = (\pi, 0)$, respectively. Past null infinity 
$\scri^-$ and future null infinity $\scri^+$ are given by the hyper-surfaces $T = R - \pi$ and $T = \pi - R$ 
with $0 < R < \pi$, respectively. 

\subsection{Blowing up of space-like infinity to a cylinder}
\label{sec:blow_up}

\begin{figure}[htb]
 \centering
  \subfigure[]{
   \includegraphics[scale = 0.57]{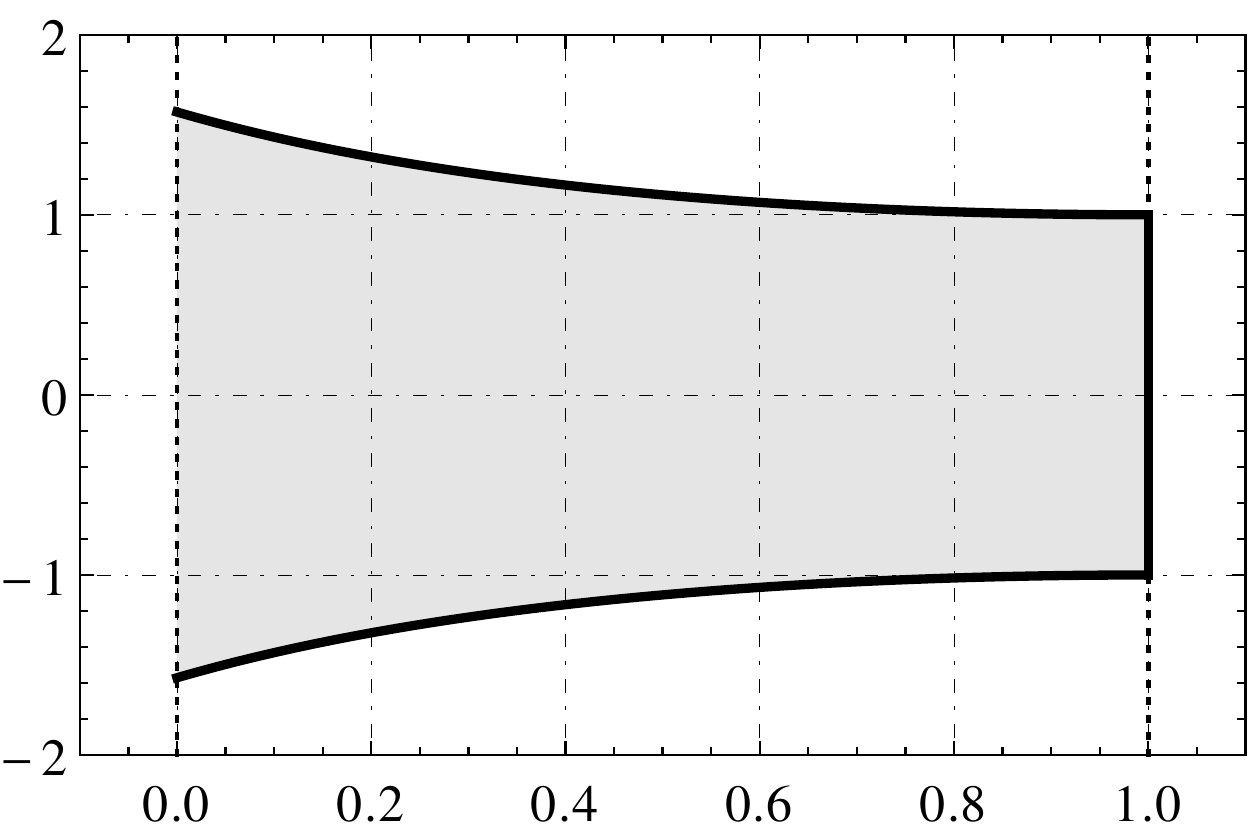}
   \label{subfig:non_horizontal_compact}
   \put(-190,22){$i^-$}
   \put(-100,30){$\scri^-$}
   \put(-18,40){$I^-$}
   \put(-15,70){$I$}
   \put(-18,100){$I^+$}
   \put(-100,110){$\scri^+$}
   \put(-190,120){$i^+$}
   \put(-215,69){\Large{$t$}}
   \put(-100,-3){\Large{$r$}}
  }
  \hspace{0.3cm}
  \subfigure[]{
   \includegraphics[scale = 0.59]{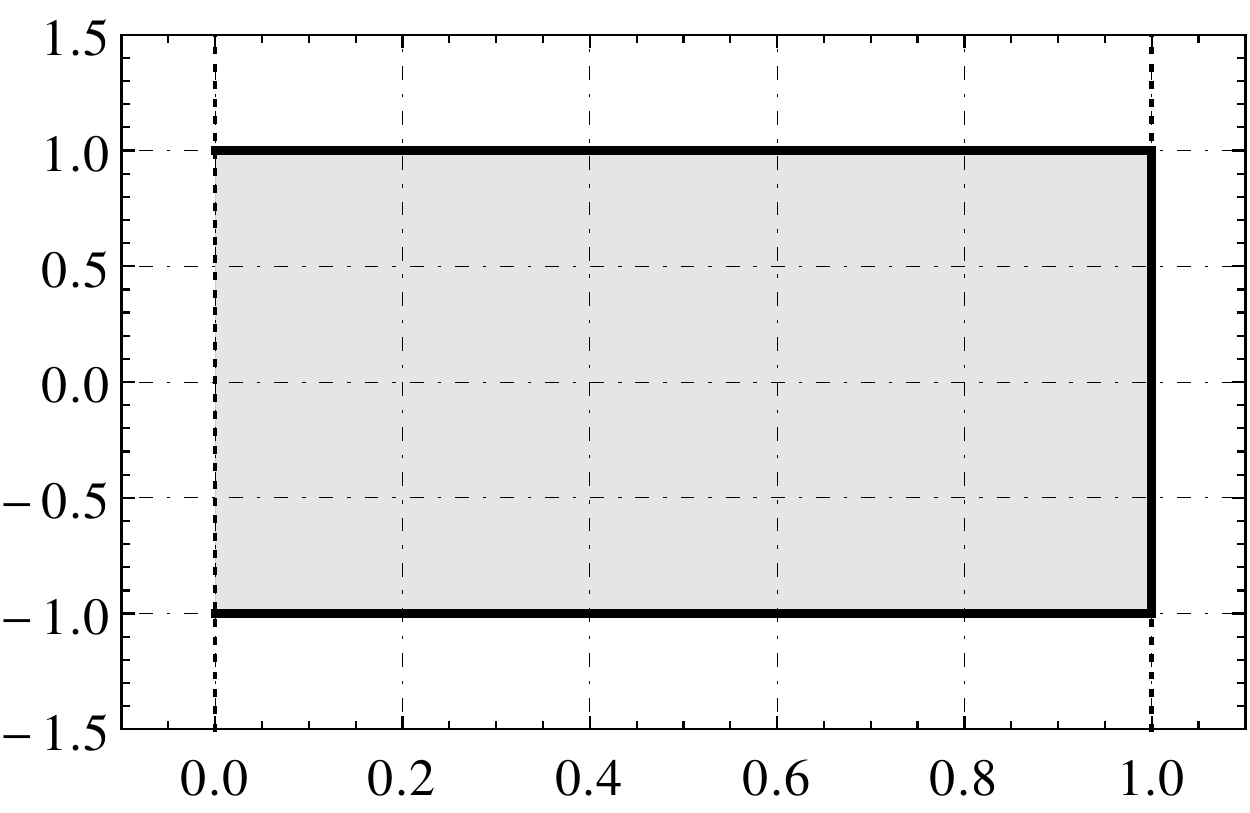}
   \label{subfig:horizontal_compact}
   \put(-190,30){$i^-$}
   \put(-110,25){$\scri^-$}
   \put(-18,30){$I^-$}
   \put(-15,70){$I$}
   \put(-18,110){$I^+$}
   \put(-110,115){$\scri^+$}
   \put(-190,110){$i^+$}
   \put(-220,69){\Large{$t$}}
   \put(-100,-3){\Large{$r$}}
  }
 \caption{Conformal compactification of Minkowski space-time into the Einstein static universe with space-like 
 infinity $i^0$ been blown up to a cylinder of finite size. The Einstein static universe $\mathbb{E}$ spans 
 the area between the dotted vertical lines at $r = 0$ and $r = 1$. Conformal diagrams of Minkowski space-time 
 embedded in the Einstein static universe \eqref{cyl_metric} on the $(t, r)$ plane for the choices $\kappa(r) = 
 \cos \left(\frac{\pi\, r}{2} \right)$ and (a) $f(t) = 2\,t$ and (b) $f(t) = \frac{1}{20}\, \mathrm{arctanh} (t)$. 
 The shaded areas denote the part of $\mathbb{E}$ that is conformal to the Minkowski space-time. The structure 
 of conformal infinity for the (a) non-horizontal \eqref{infinity_non_horizontal} and (b) horizontal \eqref{infinity_horizontal} 
 case is clearly visible.}
 \label{fig:compactification_cyl}
\end{figure}
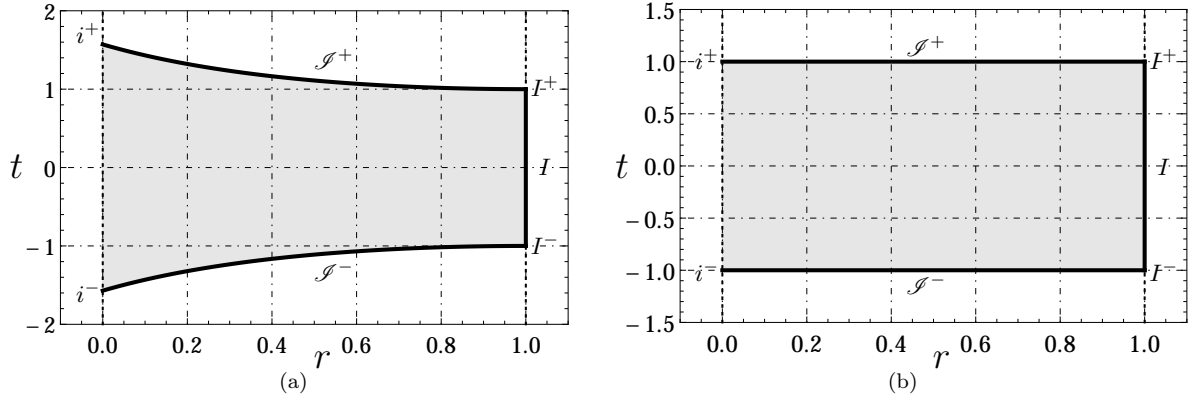

Although the metric \eqref{mink_comp_metric} extends smoothly to space-like infinity, reconstructing the
Minkowski space-time described by it, from initial data that satisfy the conformal constraints, is not 
trivial at all as some of the initial data exhibit a singular behaviour at the point $i^0$. Following the
discussion in \cite{Friedrich1998}, one can render the initial data regular by performing an appropriate 
rescaling of \eqref{mink_comp_metric}. In this new picture space-like infinity $i^0$ has a finite representation 
as a cylinder, see Fig.~\ref{fig:compactification_cyl}. In the following, we describe how to blow up the 
point $(T, R) = (0, \pi)$ that represents space-like infinity $i^0$ on the Einstein static universe.

First, in accordance with \cite{Friedrich1998}, the space-time metric \eqref{mink_comp_metric} is rescaled 
to
\begin{equation*}
 g = \kappa^{-2} g_M = \kappa^{-2} \left[ \dd T^2 - \dd R^2 - \sin^2 \!R \left(\dd \theta^2 + \sin^2 \!\theta \,\dd \phi^2 \right) \right]
\end{equation*}
and then a new time $t$ and spatial $r$ coordinate is introduced by the transformation $(T, R) = \left(\kappa(r) f(t), \pi\, r \right)$ 
to get
\begin{equation}
 \label{cyl_metric}
  g = \dot{f}^2\, \dd t^2 + \frac{2 \kappa' f \dot{f}}{\kappa}\, \dd t\, \dd r - \frac{(\pi^2 - f^2 \kappa'^2)}{\kappa^2}\, \dd r^2 -
  \frac{\sin^2(\pi\, r)}{\kappa^2} \left(\dd \theta^2 + \sin^2 \!\theta \,\dd \phi^2 \right),
\end{equation}
where $\dot{}$ and $'$ denote differentiation with respect to $t$ and $r$, respectively. Notice that 
\eqref{cyl_metric} is spherically symmetric and that the new spatial coordinate ranges in the interval 
$0 < r < 1$. As we will see below, the functions $\kappa(r)$ and $f(t)$ control the shape and the 
location of the cylinder and of null infinity $\scri^\pm$ with respect to the coordinates $t$ and $r$. 

Specifically, as we will see in the next section, the rescaling function $\kappa$ multiplies the spatial 
derivatives in the evolution equations. Thus, according to \cite{Friedrich1998}, $\kappa$ must vanish on 
the cylinder, i.e. at $r = 1$. In addition, $\kappa$ must be even with respect to $r = 0$ to maintain a 
regular centre. There is a plethora of functions that satisfy these criteria; in the following will work 
with the choice 
\begin{equation}
 \label{kappa}
  \kappa(r) = \cos \left(\frac{\pi\, r}{2} \right).
\end{equation}
The time dependent function $f$, on the other hand, appears in the coefficients of the time derivatives 
of the evolution equations, thus it controls the size of the cylinder and the shape of null infinity 
$\scri^\pm$. It turns out that the vanishing of the overall conformal factor $\Theta = \kappa^{-1} \Omega$ 
between the metrics \eqref{mink_metric} and \eqref{cyl_metric}, i.e. $g = \Theta^2\, \bar{g}_M$, dictates 
the choice of $f(t)$. Observing \eqref{Omega} and \eqref{kappa}, the conformal factor $\Theta$ in the 
new $(t, r)$ coordinates reads
\begin{equation}
 \label{Theta}
  \Theta(t, r) = \frac{\Omega}{\kappa} = \frac{1}{\kappa} \left[\cos \left(\kappa\, f \right) + \cos \left(\pi\, r\right)\right].
\end{equation}
As before, the vanishing of \eqref{Theta} locates the position of conformal infinity. Notice that now 
space-like infinity is not represented as a point but as a cylinder of finite temporal extension with 
respect to the time coordinate $t$. For $0 \leq r < 1$ the condition $\Theta = 0$ is equivalent to 
\begin{equation}
 \label{scri_t}
  t = \pm\, f^{-1} \left(\frac{\pi (1 - r)}{\kappa} \right),
\end{equation}
where $f^{-1}$ is the inverse of the time dependent function $f(t)$. The position of $\scri^\pm$ immediately 
follows 
\begin{equation}
 \label{scri_position}
  \scri^\pm = \left\{ 0 \leq r < 1, \quad t = \pm\, f^{-1} \left(\frac{\pi (1 - r)}{\kappa} \right) \right\}.
\end{equation}
In addition, at $r = 1$ the condition $\Theta = 0$ is always satisfied, indicating the presence of the cylinder 
at this location. As expected, at the limit $r \rightarrow 1$, future and past null infinity do not meet at the 
same point as they do in the conventional picture, see Fig.~\ref{fig:compactification_cyl}. (Notice that \eqref{scri_t} 
for $r \rightarrow 1$ gives $t = \pm f^{-1} (2)$.) Now, null and space-like infinity meet at the so-called 
critical sets
\begin{equation}
 \label{critical_sets}
  I^\pm = \left \{r = 1, \quad t = \pm\, f^{-1} (2) \right \}
\end{equation}
that are 2-spheres representing the bases of the cylinder. So, the sets \eqref{critical_sets} are bounding 
from above and below the cylinder and consequently its position is given by the set 
\begin{equation}
 \label{cylinder_I}
  I = \left \{r = 1, \quad - f^{-1} (2) < t < f^{-1} (2) \right \}.
\end{equation}
It is noteworthy that the height of the cylinder $I$ is $H_I = 2\, f^{-1} (2)$. 
  
In the rest of the section, we will justify the choices of the time dependent function $f$ that are going 
to be employed in the present work. One of the simplest possible choices for $f(t)$ is 
\begin{equation}
 \label{f_non_horizontal}
  f(t) = 2\, t, 
\end{equation}
whose inverse is the function $f^{-1}(y) = y/2$. Therefore, the structure of conformal infinity for the 
choice \eqref{f_non_horizontal} follows from \eqref{scri_position}-\eqref{cylinder_I}: 
\begin{equation}
 \label{infinity_non_horizontal}
 I = \left \{r = 1, \,\, - 1 < t < 1 \right \}, \quad I^\pm = \left \{r = 1, \,\, t = \pm\, 1 \right \}, 
 \quad \scri^\pm = \left\{ 0 \leq r < 1, \,\, t = \pm\, \frac{\pi (1 - r)}{2\,\kappa} \right\}.
\end{equation}
A graphic representation of \eqref{infinity_non_horizontal} is depicted in Fig.~\ref{subfig:non_horizontal_compact}. 
If one wants to make $\scri^\pm$ horizontal, then the function $f^{-1}$ in \eqref{scri_t} must be chosen 
in such a way that it is constant in its domain of definition. A possible choice of $f$ that has this 
property is 
\begin{equation}
 \label{f_horizontal}
  f(t) = \frac{1}{20}\, \mathrm{arctanh} (t). 
\end{equation}
Notice that the inverse of \eqref{f_horizontal}, i.e. $f^{-1}(y) = \tanh(20\, y)$, is constant within 
machine accuracy in its domain $D_{f^{-1}} = [ 40,\, 20 \pi ]$. Specifically, the expression \eqref{scri_t} 
for the choice \eqref{f_horizontal} reads $t = \pm 1$ and consequently conformal infinity, see Fig.~\ref{subfig:horizontal_compact}, 
has the following structure  
\begin{equation}
 \label{infinity_horizontal}
 I = \left \{r = 1, \,\, - 1 < t < 1 \right \}, \quad I^\pm = \left \{r = 1, \,\, t = \pm\, 1 \right \}, 
 \quad \scri^\pm = \left\{ 0 \leq r < 1, \,\, t = \pm\, 1 \right\}.
\end{equation}
The main advantage of the horizontal representation \eqref{infinity_horizontal} is that the whole of Minkowski 
space-time has been mapped to a rectangle of finite size, see Fig.~\ref{subfig:horizontal_compact}. As we will 
see in Sec.~\ref{sec:numerics}, this representation is numerically very advantageous.   


\section{Generalised conformal field equations}
\label{sec:GCFE}

Here the generalised conformal field equations \cite{Friedrich1998} will be linearised on Minkowski space-time.
On a Minkowski background, as the one described in the previous section, their linearisation simplifies 
considerably. Namely, the so-called structural equations defining the torsion and the curvature are trivially 
satisfied \cite{Friedrich2003,DoulisPhD}, thus one is left with the remaining Bianchi identity for the 
perturbation of the rescaled Weyl tensor. In the present section, the spinor version of this equation, 
i.e. the so-called spin-2 zero-rest-mass equation \cite{Penrose1965}, will be studied analytically as a 
system of first and second order partial differential equations (PDEs). Therefore, in the rest of this 
work the 2-spinor formalism \cite{Pen&Rind1986} will be used. We employ this formalism as it simplifies 
our arguments and makes the manipulation of the quantities relevant to our study---which also emerge more 
naturally in this formalism---easier.

\subsection{Spin-2 zero-rest-mass equation}
\label{sec:spin2}

Here, we describe how to obtain a coordinate representation of the spin-2 zero-rest-mass equation as a 
system of first order PDEs and study thoroughly its analytical properties. 

\subsubsection{Basis, connections, and spin-coefficients}

We start by defining a basis and a connection compatible with \eqref{cyl_metric}. To do so, we first 
introduce a spin-frame $(o^A, \iota^A)$, with the usual normalization condition $o_A\, \iota^A = 1$, and 
then form a coordinate basis  by defining the non-orthonormal null tetrad $(l^\mu , n^\mu , m^\mu , \bar{m}^\mu)$ 
with
\begin{align}
 \begin{split}
  \label{null_tetrad}
   l^\mu = \frac{1}{\sqrt{2}} \left(A, B, 0, 0 \right),& \quad 
   n^\mu = \frac{1}{\sqrt{2}} \left(C, -B, 0, 0 \right), \\
   m^\mu = \frac{1}{\sqrt{2}\, g\, r} \left(0, 0, 1, - \mathrm{i}\, \csc \theta \right),& \quad
   \bar{m}^\mu = \frac{1}{\sqrt{2}\, g\, r} \left(0, 0, 1, \mathrm{i}\, \csc \theta \right),
 \end{split}
\end{align}
where the functions $A(t, r), B(t, r)$ $C(t, r)$, and $g(r)$ are uniquely defined in terms of the metric 
coefficients of \eqref{cyl_metric}, i.e.
\begin{equation}
 \label{metric_functions}
  A = \frac{1}{\dot{f}} \left(1 - \frac{\kappa' f}{\pi} \right), \qquad 
  B = \frac{\kappa}{\pi}, \qquad
  C = \frac{1}{\dot{f}} \left(1 + \frac{\kappa' f}{\pi} \right), \qquad
  g = \frac{\sin(\pi\, r)}{\kappa\, r}.
\end{equation}
It can be readily confirmed that the null vectors \eqref{null_tetrad} satisfy the correct inner product 
conditions: $l^\mu n_\mu = - m^\mu \bar{m}_\mu = 1$ and all the remaining combinations vanish.

The directional derivatives along the null vectors \eqref{null_tetrad} immediately follow
\begin{align}
 \begin{split}
  \label{intrinsic_derivatives}
   D = l^\mu \partial_\mu = \frac{1}{\sqrt{2}} \left(A\, \partial_t + B\, \partial_r \right),& \quad 
   D' = n^\mu \partial_\mu = \frac{1}{\sqrt{2}} \left(C\, \partial_t - B\, \partial_r \right), \\
   \delta = m^\mu \partial_\mu = \frac{1}{\sqrt{2}\, g\, r} \left(\partial_\theta - \mathrm{i}\, \csc \theta\, \partial_\phi \right),& \quad
   \delta' = \bar{m}^\mu \partial_\mu = \frac{1}{\sqrt{2}\, g\, r} \left(\partial_\theta + \mathrm{i}\, \csc \theta\, \partial_\phi \right),
 \end{split}
\end{align}
where $A, B, C, g$ are as above. The commutation relations among the directional derivatives \eqref{intrinsic_derivatives} 
provide us with the  spin-coefficients corresponding to \eqref{cyl_metric} and \eqref{null_tetrad}. The 
non-vanishing spin-coefficients are listed below
\begin{equation}
 \label{spin_coef}
  \alpha = - \beta = \frac{\cot \theta}{2 \sqrt{2}\, g\, r}, \quad 
  \gamma = \epsilon = -\frac{\kappa'}{2 \sqrt{2}\, \pi}, \quad 
  \rho = - \rho' = \frac{\kappa' - \pi\,  \kappa \cot (\pi\, r)}{\sqrt{2}\, \pi}.
\end{equation}
Next, we define the weighted differential operators of the GHP (Geroch-Held-Penrose) formalism in terms 
of the directional derivatives \eqref{intrinsic_derivatives} and the spin-coefficients \eqref{spin_coef}:
\begin{align}
 \begin{split}
  \label{GHP_derivatives}
   \mbox{\th}\, \eta = \left(D - 2\, w\, \gamma \right) \eta,& \quad 
   \mbox{\th}' \eta = \left(D' - 2\, w\, \gamma \right) \eta, \\
   \eth\, \eta = \left(\delta + 2\, s\, \alpha \right) \eta,& \quad 
   \eth' \eta = \left(\delta' - 2\, s\, \alpha \right) \eta,
 \end{split}
\end{align}
where $\eta$ is a $\{p, q\}$-scalar quantity with boost-weight $w = \frac{p+q}{2}$ and spin-weight 
$s = \frac{p-q}2$, see~\cite{Pen&Rind1986}.

\subsubsection{Derivation}
\label{sec:spin2_derivation}

Having defined a basis, derivatives, and spin-coefficients compatible with the metric \eqref{cyl_metric}, 
we proceed further and express the spin-2 zero-rest-mass equation as a system of first order PDEs. Our 
starting point is the spinor version of the Bianchi identity for the Weyl tensor on a general space-time 
with metric $\tilde g$, which in the absence of matter reads 
 \begin{equation}
 \label{spinor_Bianchi_Weyl}
  \widetilde\nabla^A{}_{A'} \widetilde\Psi_{ABCD} = 0,
\end{equation}
where $\widetilde\Psi_{ABCD}$ is the spinor counterpart of the Weyl tensor for the metric
$\tilde g$---$\Psi_{ABCD}$ is called the Weyl spinor and is totally symmetric in its indices
$\widetilde\Psi_{ABCD} = \widetilde\Psi_{(ABCD)}$---and $\widetilde\nabla_{AA'}$ is the spinor 
covariant derivative, see \cite{Pen&Rind1986}. The generalised conformal field equations are concerned
with the conformal structure of a space-time, i.e., the structure that remains invariant under
conformal rescalings of the metric $\tilde g \mapsto g = \Theta^2 \tilde g$. Under these
transformations the Weyl spinor remains invariant $\Psi_{ABCD} = \widetilde\Psi_{ABCD}$. The
behaviour of the spinor Bianchi identity \eqref{spinor_Bianchi_Weyl} under this kind of conformal
transformations reads \cite{Pen&Rind1986}
\begin{equation*}
 \nabla^A{}_{A'} \left(\Theta^{-1}\Psi_{ABCD}\right) = \Theta^{-2}\, \widetilde\nabla^A{}_{A'} \widetilde\Psi_{ABCD}, 
\end{equation*}
which through \eqref{spinor_Bianchi_Weyl} results in
\begin{equation}
 \label{spinor_Bianchi_Weyl_rescaled}
  \nabla^A{}_{A'} \Delta_{ABCD} = 0,
\end{equation}
the so-called Bianchi equation for the rescaled Weyl spinor $\Delta_{ABCD} = \Theta^{-1} \Psi_{ABCD}$.

Now, in order to obtain the weak field limit of \eqref{spinor_Bianchi_Weyl_rescaled}, i.e. the equation 
for a small perturbation of a Minkowski space-time, we consider, as in \cite{Pen&Rind1986}, that our 
space-time metric is a smoothly varying function $g(u)$ of a single parameter $u$ such that $g(0)$ is 
conformal to the Minkowski space-time. Similarly, one expects that the rescaled Weyl spinor $\Delta_{ABCD}(u)$ 
for any non-zero value of $u$ satisfies \eqref{spinor_Bianchi_Weyl_rescaled} and tends smoothly to zero 
as $u \rightarrow 0$. The latter guarantees that the totally symmetric spinor quantity $u^{-1} \Delta_{ABCD}(u)$ 
has a well defined limit, say $\Phi_{ABCD} $, as $u \rightarrow 0$ and satisfies the equation
\begin{equation}
 \label{spin2_equation}
  \nabla^A{}_{A'} \Phi_{ABCD} = 0,
\end{equation}
where the totally symmetric spinor $\Phi_{ABCD}$ will be called the spin-2 zero-rest-mass field. The spin-2 
zero-rest-mass equation \eqref{spin2_equation} is the spinor version of the Bianchi equation for the rescaled 
Weyl tensor.

\subsubsection{Coordinate representation}

In order to proceed further in our study of \eqref{spin2_equation}, we have to decompose it in its components. 
To do so, one has first to express the spinor covariant derivative in \eqref{spin2_equation}, as in \cite{DoulisPhD}, 
in terms of the weighted differential operators \eqref{GHP_derivatives} of the GHP formalism:
\begin{equation}
 \label{covariant_deriv}
  \nabla^A{}_{A'} = \iota^A \iota_{A'} \mbox{\th} + o^A o_{A'} \mbox{\th}' - \iota^A o_{A'} \eth - o^A \iota_{A'} \eth'; 
\end{equation}
then substitute this expression in \eqref{spin2_equation} and expand the spin-2 field in terms of its 
five independent components $(\Phi_0, \Phi_1, \Phi_2, \Phi_3, \Phi_4)$; and finally take the components 
of the resulting expression to get
\begin{align}
 \begin{split}
  \label{spin2_equation_GHP}
   & \mbox{\th}\, \Phi_k - \eth' \Phi_{k-1} = (5-k)\, \rho\, \Phi_k, \\
   & \mbox{\th}' \Phi_{k-1}  - \eth\, \Phi_k = -k\, \rho\, \Phi_{k-1},
 \end{split}
\end{align}
where $k = 1,2,3,4$ and $\rho$ is given by \eqref{spin_coef}. Next, by using \eqref{GHP_derivatives} and 
\eqref{intrinsic_derivatives} it is possible to obtain a coordinate representation of the system \eqref{spin2_equation_GHP} 
\begin{align}
 \begin{split}
  \label{spin2_equation_coord}
   & A\, \partial_t \Phi_k + B\, \partial_r \Phi_k - \sqrt{2}\, \left[(5 - k)\, \rho + 2\,(2 - k)\, \epsilon \right] \Phi_k = \sqrt{2}\,\, \eth'\, \Phi_{k-1}, \\
   & C\, \partial_t \Phi_{k-1} - B\, \partial_r \Phi_{k-1} + \sqrt{2}\, \left[k\, \rho - 2\,(3 - k)\, \epsilon \right] \Phi_{k-1} = \sqrt{2}\,\, \eth\, \Phi_k,
 \end{split}
\end{align}
where $k = 1,2,3,4$, the function $A, B, C$ are given by \eqref{metric_functions}, and the spin-coefficients 
$\rho, \epsilon$ by \eqref{spin_coef}. Observing \eqref{intrinsic_derivatives}-\eqref{GHP_derivatives}, 
the $\eth, \eth'$ operators can be expressed on the unit sphere through the transition
\begin{equation}
 \label{eth_unit}
  \eth \mapsto \frac{1}{\sqrt{2}\, g\, r} \eth_0 \quad \mbox{and} \quad \eth' \mapsto \frac{1}{\sqrt{2}\, g\, r} \eth'_0, 
\end{equation}
where $\eth_0 = \partial_\theta - \mathrm{i}\, \csc \theta\, \partial_\phi + s\, \cot \theta$ and $\eth'_0 = 
\partial_\theta + \mathrm{i}\, \csc \theta\, \partial_\phi - s\, \cot \theta$ denote the ``eth'' operators 
on the unit sphere. Finally, we use the spherical symmetry of the metric \eqref{cyl_metric} to expand 
the components $\phi_k$ of the spin-2 field as a sum of spin-weighted spherical harmonics ${}_s Y_{lm}$ 
in the following way 
\begin{equation}
 \label{spher_expansion}
  \phi_k(t, r, \theta, \phi) = \sum_{lm} \phi_k^{lm}(t,r)\; {}_{2-k} Y_{lm}(\theta, \phi),
\end{equation}
where $s = 2 - k$ is the spin-weight of $\phi_k$ and the integers $s, l, m$ satisfy the inequalities 
$|s|\leq l$ and $|m|\leq l$. Since the operators $\eth_0, \eth_0'$ act on the spin-weighted spherical 
harmonics $_s Y_{lm}$ as
\begin{equation}
  \begin{aligned}
   \label{eth_on_spher_harm}
    \eth_0(_s Y_{lm}) &= - \sqrt{l(l+1)-s(s+1)}\; {}_{s+1} Y_{lm}, \\
    \eth_0'(_s Y_{lm}) &= \sqrt{l(l+1)-s(s-1)}\;  {}_{s-1} Y_{lm},
  \end{aligned}
\end{equation} 
the system \eqref{spin2_equation_coord} decouples into separate systems for each mode of the fixed pair 
$(l, m)$, i.e.
\begin{align}
 \begin{split}
  \label{spin2_equation_final}
   & A\, \partial_t \Phi_k + B\, \partial_r \Phi_k - \sqrt{2}\, \left[(5 - k)\, \rho + 2\,(2 - k)\, \epsilon \right] \Phi_k = \frac{\alpha_{(2-k)(3-k)}}{g\, r}\, \Phi_{k-1}, \\
   & C\, \partial_t \Phi_{k-1} - B\, \partial_r \Phi_{k-1} + \sqrt{2}\, \left[k\, \rho - 2\,(3 - k)\, \epsilon \right] \Phi_{k-1} = -\frac{\alpha_{(2-k)(3-k)}}{g\, r}\, \Phi_k,
 \end{split}
\end{align}
where $k = 1,2,3,4$, $\alpha_n \equiv \sqrt{l(l+1) - n}$, the function $A,B,C,g$ are given by \eqref{metric_functions}, 
and the spin-coefficients $\rho, \epsilon$ by \eqref{spin_coef}. The eight equations \eqref{spin2_equation_final}, 
for the five independent components $\Phi_k$ of the spin-2 field, comprise a coordinate representation 
of \eqref{spin2_equation} on the background space-time \eqref{cyl_metric}. 

\subsubsection{Evolution and constraint equations}
\label{sec:spin2_evolution}

The system \eqref{spin2_equation_final} can be readily split into five evolution equations
\begin{align}
 \begin{split}
  \label{spin2_equation_evol}
   C\, \partial_t \Phi_0 &= B\, \partial_r \Phi_0 + 4 \sqrt{2}\, \epsilon\,  \Phi_0 - \sqrt{2}\, \rho\,  \Phi_0 - \frac{\alpha_2}{g\, r}\, \Phi_1, \\
   (A + C)\, \partial_t \Phi_1 &= 4 \sqrt{2}\, \epsilon\,  \Phi_1 + 2 \sqrt{2}\, \rho\,  \Phi_1 + \frac{\alpha_2}{g\, r}\, \Phi_0 - \frac{\alpha_0}{g\, r}\, \Phi_2, \\
   (A + C)\, \partial_t \Phi_2 &= \frac{\alpha_0}{g\, r}\, \Phi_1 - \frac{\alpha_0}{g\, r}\, \Phi_3, \\
   (A + C)\, \partial_t \Phi_3 &= -4 \sqrt{2}\, \epsilon\,  \Phi_3 - 2 \sqrt{2}\, \rho\,  \Phi_3 + \frac{\alpha_0}{g\, r}\, \Phi_2 - \frac{\alpha_2}{g\, r}\, \Phi_4, \\
   A\, \partial_t \Phi_4 &= -B\, \partial_r \Phi_4 - 4 \sqrt{2}\, \epsilon\,  \Phi_4 + \sqrt{2}\, \rho\,  \Phi_4 + \frac{\alpha_2}{g\, r}\, \Phi_3
 \end{split}
\end{align}
and three constraints
\begin{eqnarray}
  \label{spin2_equation_constr}
   && \nonumber C_1 \equiv 2\, (A + C)\, B\, \partial_r \Phi_1 + 4\, \sqrt{2}\, A\, (\epsilon - \rho)\, \Phi_1 - 4\, \sqrt{2}\, C\, (2\, \rho + \epsilon)\, \Phi_1 - \
   \frac{2\, \alpha_0}{g\, r}\, A\, \Phi_2 - \frac{2\, \alpha_2}{g\, r}\, C\, \Phi_0 = 0, \\
   && C_2 \equiv 2\, (A + C)\, B\, \partial_r \Phi_2 - 6\, \sqrt{2}\, (A + C)\, \rho\,  \Phi_2 - \frac{2\, \alpha_0}{g\, r}\, C\, \Phi_1 - \frac{2\, \alpha_0}{g\, r}\, A\, \Phi_3 = 0, \\
   && \nonumber C_3 \equiv 2\, (A + C)\, B\, \partial_r \Phi_3 + 4\, \sqrt{2}\, C\, (\epsilon - \rho)\, \Phi_3 - 4\, \sqrt{2}\, A\, (2\, \rho + \epsilon)\, \Phi_3 - \
   \frac{2\, \alpha_0}{g\, r}\, C\, \Phi_2 - \frac{2\, \alpha_2}{g\, r}\, A\, \Phi_4 = 0. 
\end{eqnarray}
Notice that the evolution system \eqref{spin2_equation_evol} acquires a very simple form on the cylinder 
$I$ (i.e. at $r = 1$). The radial derivatives are multiplied by the function $B$ defined in \eqref{metric_functions}, 
which for the choice \eqref{kappa} vanishes at $r = 1$. Thus, the radial derivatives drop out from the
equations controlling the dynamics of the components $\Phi_0, \Phi_4$ and the cylinder $I$ becomes a total 
characteristic of the system \eqref{spin2_equation_evol}. This feature, together with the fact that $I$, 
as can be seen in Fig.~\ref{fig:compactification_cyl}, lies on the boundary located at $r = 1$ of the 
computational domain, indicates that we are not allowed to impose boundary conditions at points lying on 
$I$. 

To check the well-posedness of the Cauchy problem described by the system \eqref{spin2_equation_evol}-\eqref{spin2_equation_constr}, 
one has to write the evolution equations in a matrix form for the vector $\mathbf{\Phi} \equiv (\Phi_0, \Phi_1, \Phi_2, \Phi_3, \Phi_4)^\mathrm{T}$:
\begin{equation*}
 \mathbf{A}_0 \partial_t \mathbf{\Phi} + \mathbf{A}_1 \partial_r \mathbf{\Phi} = \mathbf{A}_2\, \mathbf{\Phi},
\end{equation*}
where $\mathbf{A}_0 = \mathrm{diag}(A, A + C, A + C, A + C, C)$ and $\mathbf{A}_1 = \mathrm{diag}(-B, 0, 0, 0, B)$ 
are obviously Hermitian matrices and $\mathbf{A}^0$ is positive definite when the conditions $\dot{f}(t) > 0$ 
and $|f| < \pi/|\kappa'|$ hold simultaneously. Therefore, the evolution system is symmetric hyperbolic, and 
consequently well-posed, in the range $|t| < f^{-1}(\pi/|\kappa'|)$---given that the condition $\dot{f} > 0$ 
is also satisfied in this range. 

To prove that the constraints \eqref{spin2_equation_constr} are preserved during the evolution, one has 
to derive the subsidiary system for the constraint quantities $C_k$ appearing in \eqref{spin2_equation_constr}. 
\begin{align*}
 \begin{split}
  & \partial_t C_1 = - \kappa\, \dot{f}\, \cot(\pi\, r)\, C_1 - \frac{1}{2}\, \alpha_0\, \kappa\, \dot{f}\, \csc(\pi  r)\, C_2, \\
  & \partial_t C_2 = \frac{1}{2}\, \alpha_0\, \kappa\, \dot{f}\, \csc(\pi\, r)\, C_1 - \frac{1}{2}\, \alpha_0\, \kappa\, \dot{f}\, \csc(\pi\, r)\, C_3, \\ 
  & \partial_t C_3 = \frac{1}{2}\, \alpha_0\, \kappa\, \dot{f}\, \csc(\pi\, r)\, C_2 + \kappa\, \dot{f}\, \cot(\pi\, r) C_3, 
 \end{split}
\end{align*}
where, for the sake of simplicity, the functions $\kappa, f$ were used here instead of the $A, B, C, g$ 
functions and the spin-coefficients $\rho, \epsilon$. The above system can be written in the symmetric
hyperbolic form $\partial_t \mathbf{C} = \mathbf{A}_3 \mathbf{C}$ for the vector $\mathbf{C} \equiv (C_1, C_2, C_3)^\mathrm{T}$, 
where the entries of the $3\times3$ matrix $\mathbf{A}_3$ consist of the coefficients of the r.h.s of 
the above subsidiary system. The eigenvalues of $\mathbf{A}_3$ read $\lambda = 0, \pm \mathrm{i}\, \csc(\pi r)\, 
\kappa\, \dot{f}\, \sqrt{\alpha_0^2 - 2 \cos^2(\pi r)}/\sqrt{2}$. The imaginary nature of the non-trivial 
eigenvalues guarantees that the unavoidable initial violation of the constraints does not grow exponentially 
during the evolution.

\subsubsection{Characteristic curves}

\begin{figure}[htb]
 \centering
  \subfigure[]{
   \includegraphics[scale = 0.56]{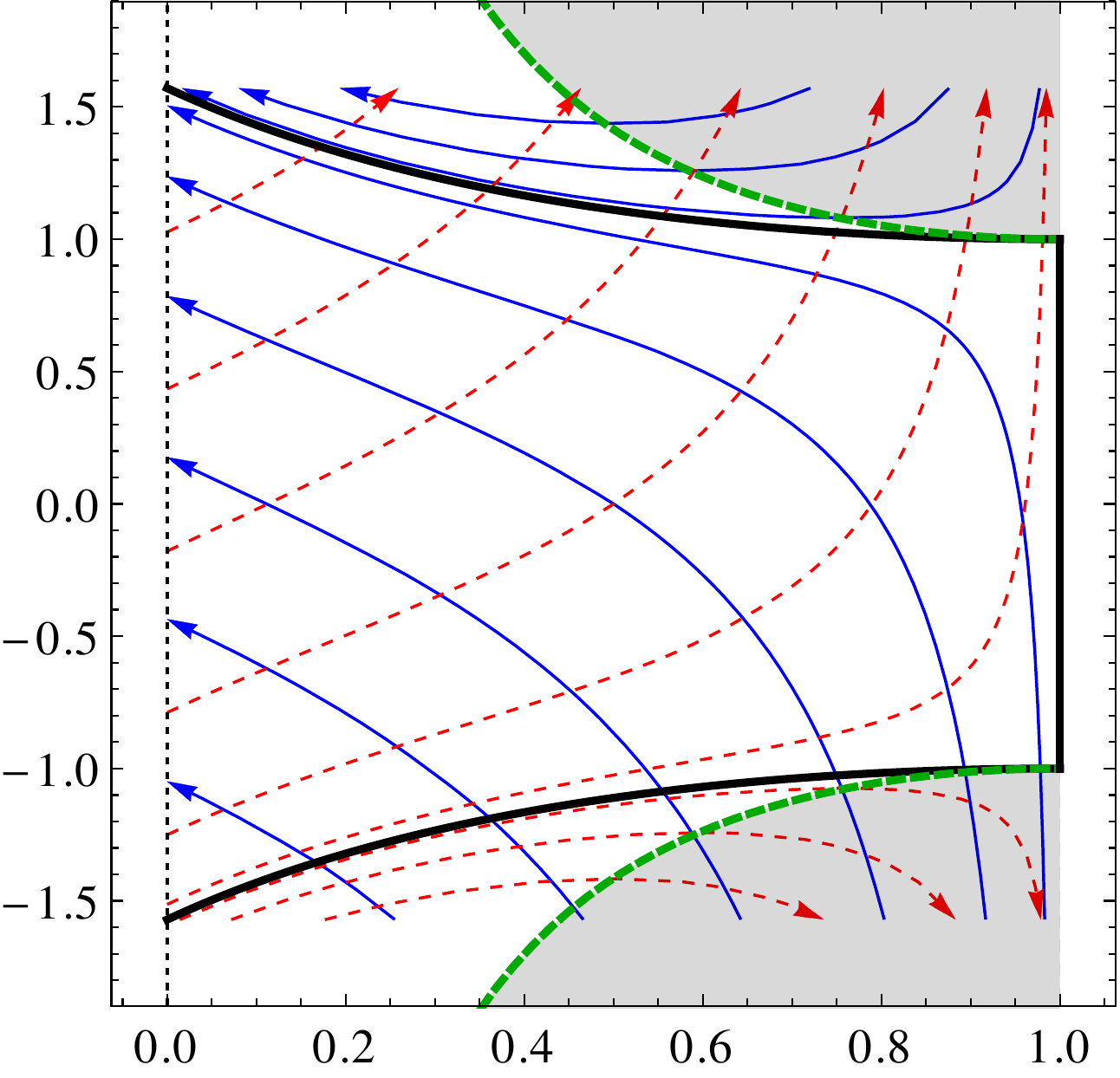}
   \label{subfig:non_horizontal_charact}
   \put(-190,20){$i^-$}
   \put(-105,43){$\scri^-$}
   \put(-16,48){$I^-$}
   \put(-11,105){$I$}
   \put(-14,160){$I^+$}
   \put(-105,165){$\scri^+$}
   \put(-188,185){$i^+$}
   \put(-215,105){\Large{$t$}}
   \put(-100,-3){\Large{$r$}}
  }
  \hspace{0.3cm}
  \subfigure[]{
   \includegraphics[scale = 0.58]{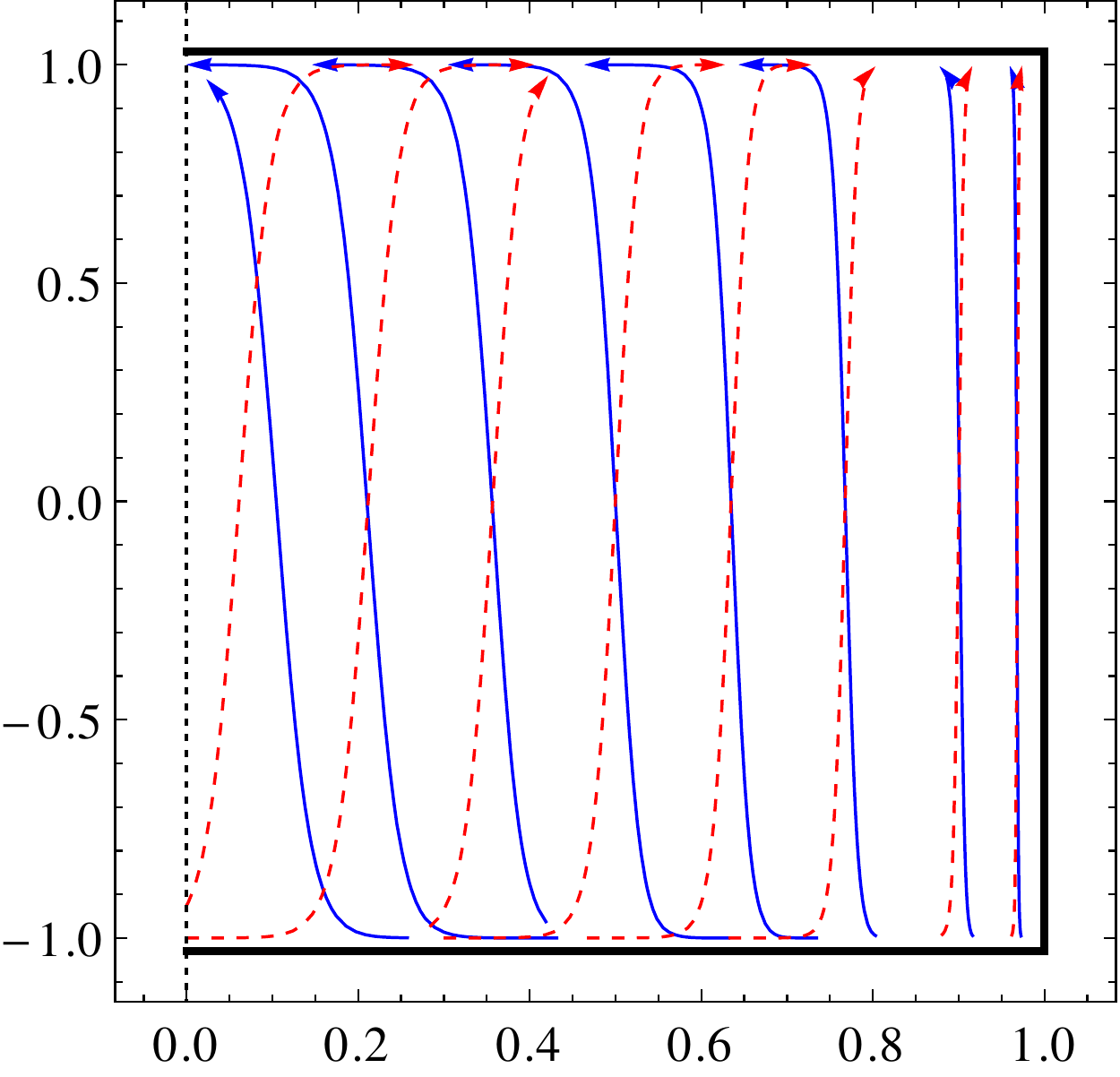}
   \label{subfig:horizontal_charact}
   \put(-188,17){$i^-$}
   \put(-110,15){$\scri^-$}
   \put(-16,20){$I^-$}
   \put(-13,105){$I$}
   \put(-15,190){$I^+$}
   \put(-110,193){$\scri^+$}
   \put(-188,190){$i^+$}
   \put(-220,105){\Large{$t$}}
   \put(-100,-3){\Large{$r$}}
  }
 \caption{Characteristic curves of the fields $\Phi_0$ (blue solid lines) and $\Phi_4$ (red dashed lines). 
 The characteristic curves for the choices (a) $f(t) = 2\,t$ and (b) $f(t) = \frac{1}{20}\, \mathrm{arctanh}(t)$ 
 are presented. The black thick solid lines represent conformal infinity and the green thick dashed lines 
 the span of the symmetric hyperbolic region. The shaded areas mark the domain of non-hyperbolicity of 
 \eqref{spin2_equation_evol}. In (b) conformal infinity and the boundary of the symmetric hyperbolic region 
 coincide, thus the system \eqref{spin2_equation_evol} is not symmetric hyperbolic in the region exterior 
 to conformal infinity (this area has not been shaded for presentational reason).}
 \label{fig:characteristics_spin2}
\end{figure}
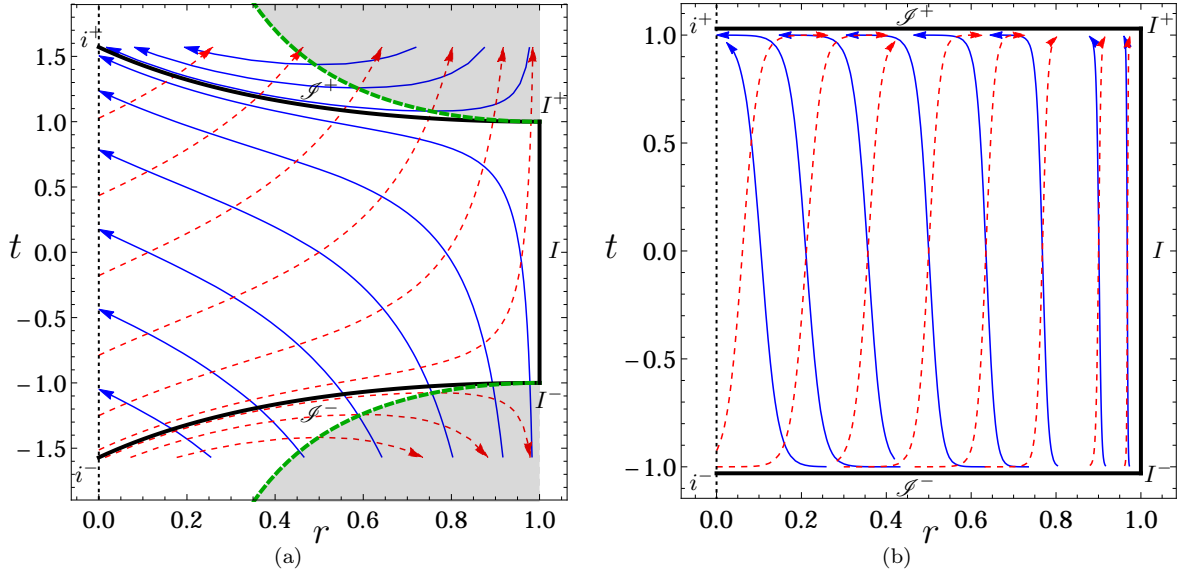

It is extremely useful to study the behaviour of the characteristic curves of the evolution system 
\eqref{spin2_equation_evol} as their form provides a qualitative insight on the behaviour of the solutions 
of \eqref{spin2_equation_evol}. For first order PDEs with principal part of the form $a(t,r)\, \partial_t u(t,r) + 
b(t,r)\, \partial_r u(t,r)$, like the ones in \eqref{spin2_equation_evol}, the slope of the characteristics
is given by $\dd t/\dd r = a/b$. For the components $\Phi_1, \Phi_2, \Phi_3$ of the spin-2 field the 
characteristics are straight lines of constant $r$ as $b$ vanishes in the corresponding evolution equations. 
The slope of the remaining two components $\Phi_0$ and $\Phi_4$ reads
\begin{equation*}
 \frac{\dd t}{\dd r} = - \frac{C}{B} \quad \mbox{and} \quad \frac{\dd t}{\dd r} = \frac{A}{B}
\end{equation*}
respectively. As expected the behaviour of the above characteristics depends entirely on the choice of 
the functions $\kappa$ and $f$, see \eqref{metric_functions}. Fig.~\ref{fig:characteristics_spin2} depicts 
the form of the characteristic curves for $\Phi_0$ and $\Phi_4$. Therein, the characteristics for the 
choices \eqref{f_non_horizontal} and \eqref{f_horizontal} of the time dependent function $f$ are presented.

Specifically, Fig.~\ref{subfig:non_horizontal_charact} depicts the characteristic curves of $\Phi_0$ 
(blue solid lines) and $\Phi_4$ (red dashed lines) for the choice \eqref{f_non_horizontal}. Notice that 
$\scri^+$ (upper black thick solid line) confines from above the characteristics of $\Phi_0$, while $\scri^-$ 
(lower black thick solid line) bounds from below the characteristics of $\Phi_4$. This observation is 
related to the well know fact \cite{Friedrich1998} that the evolution equations of $\Phi_0$ and $\Phi_4$ 
degenerate at the corresponding critical sets $I^+$ and $I^-$. The reason for this expected behaviour 
of the evolution equations is clearly visible on Fig.~\ref{subfig:non_horizontal_charact}. Namely, by 
trying to go beyond the critical sets $I^\pm$ one enters the domain of non-hyperbolicity of the equations: 
shaded areas bounded by the green thick dashed lines. In the remaining non-shaded area the system \eqref{spin2_equation_evol} 
is symmetric hyperbolic---this is the domain for which the condition of symmetric hyperbolicity $|t| < f^{-1}(\pi/|\kappa'|)$ 
is satisfied. Therefore, the whole of the conformally compactified Minkowski space-time, see shaded region 
of Fig.~\ref{subfig:non_horizontal_compact} and compare with Fig.~\ref{subfig:non_horizontal_charact}, 
is situated entirely into the symmetric hyperbolic region of \eqref{spin2_equation_evol}. As was also 
mentioned above, the shape of the characteristic curves clearly shows that the cylinder $I$ is a total 
characteristic of our system.

The situation is quite similar for the choice \eqref{f_horizontal} illustrated in Fig.~\ref{subfig:horizontal_charact}. 
The parts of $\mathbb{E}$ that are conformal to Minkowski space-time are located entirely inside the 
domain $|t| < 1$ of symmetric hyperbolicity, which now coincides with the interior of the black thick solid 
lines representing conformal infinity. Thus, there is no meaning of going beyond the critical sets $I^\pm$ 
and enter the region of non-hyperbolicity as now by reaching $I^\pm$ one actually reaches time-like infinity 
$i^\pm$. The behaviour of the characteristics here is more universal in the sense that the characteristics 
for both fields $\Phi_0$ and $\Phi_4$ are bounded from above and below by null infinity. Thus, the evolution 
equations for $\Phi_0$ and $\Phi_4$ degenerate at both critical sets $I^\pm$.

\subsection{Spin-2 zero-rest-mass wave equation}
\label{sec:spin2_wave}

In the current section, we present a coordinate representation of the spin-2 zero-rest-mass equation as 
a system of second order PDEs and discuss its analytical properties.

\subsubsection{Derivation}
\label{sec:spin2_wave_derivation}

Our starting point is the spin-2 zero-rest-mass equation \eqref{spin2_equation}. We apply the spinor 
covariant derivative $\nabla^{A'}{}_{A}$ to it and contract over the $A'$ index to get
\begin{equation*}
 \nabla_{A'A} \nabla_{F}{}^{A'} \Phi^F{}_{BCD} = 0,
\end{equation*}
where the indices have been moved appropriately in order to be able to split the above differential operator 
into its symmetric and skew symmetric parts in $AF$ 
\begin{equation*}
 \nabla_{A'A} \nabla_{F}{}^{A'} \Phi^F{}_{BCD} = \Box_{AF} \Phi^F{}_{BCD} - \frac{1}{2}\, \Box \Phi_{ABCD} = 0,
\end{equation*}
where $\Box_{AF} \equiv \nabla_{A'(A} \nabla_{F)}{}^{A'}$ and $\Box \equiv \nabla_{AA'} \nabla^{AA'}$. 
Next, using the identities 
\begin{eqnarray*}
 && \Box_{AF} k^C = \left[\Psi_{AFE}{}^C + \frac{R}{24} \left( \epsilon_{AE}\, \epsilon_{F}{}^{C} + \epsilon_{FE}\, \epsilon_{A}{}^{C} \right) \right] k^E, \\
 && \Box_{AF} k_C = -\left[\Psi_{AFC}{}^E + \frac{R}{24} \left( \epsilon_{AC}\, \epsilon_{F}{}^{E} + \epsilon_{FC}\, \epsilon_{A}{}^{E} \right) \right] k_E,
\end{eqnarray*}
where $R$ is the scalar curvature and $\Psi_{ABCD}$ is the Weyl spinor of \eqref{cyl_metric}, the above 
expression simplifies considerably
\[
 - \frac{R}{4}\, \Phi_{ABCD} + 3\, \Psi_{EF(AB} \Phi_{CD)}{}^{EF} - \frac{1}{2}\, \Box \Phi_{ABCD} = 0.
\]
Now, taking into consideration that for the metric \eqref{cyl_metric} the Weyl spinor vanishes and the 
scalar curvature reads 
\begin{equation}
 \label{ricci_scalar}
  R = \frac{3}{2}\, \kappa^2,
\end{equation}
we finally arrive at the so-called spin-2 zero-rest-mass wave equation
\begin{equation}
 \label{spin2_wave_equation}
   \Box \Phi_{ABCD} + \frac{3\, \kappa^2}{4}\, \Phi_{ABCD} = 0, 
\end{equation}
which will be used to describe \eqref{spin2_equation} as a system of second order PDEs. 

\subsubsection{Coordinate representation}

Let's obtain now a coordinate representation of \eqref{spin2_wave_equation}. As before, we express the 
differential operator $\Box$ in terms of the weighted differential operators \eqref{GHP_derivatives} of 
the GHP formalism:
\[
 \Box = 2\, \rho\, (\mbox{\th} - \mbox{\th}') + \mbox{\th} \mbox{\th}' +\mbox{\th}' \mbox{\th} - \eth \eth' - \eth' \eth; 
\]
substitute this expression into \eqref{spin2_wave_equation}; expand the spin-2 field in terms of its 
components; and take the components of the resulting expression to obtain the system of five equations 
\begin{align}
 \begin{split}
  \label{spin2_wave_equation_GHP}
   \mbox{\th} \mbox{\th}' \Phi_\lambda + \mbox{\th}' \mbox{\th} \Phi_\lambda +2\, \rho\, (\mbox{\th} - \mbox{\th}') \Phi_\lambda -\
   2\, (\lambda^2 - 4\, \lambda - 2)\, \rho^2\, \Phi_\lambda + \frac{3}{4}\, \kappa^2 \Phi_\lambda =&\\ 
   = \eth \eth' \Phi_\lambda + \eth' \eth \Phi_\lambda + 2\, (4 -& \lambda)\, \rho\, \eth \Phi_{\lambda + 1} - \lambda\, \rho\, \eth' \Phi_{\lambda - 1},
 \end{split}
\end{align}
where $\lambda = 0,1,2,3,4$, $\kappa$ is the rescaling function \eqref{kappa}, and $\rho$ is given by 
\eqref{spin_coef}. In a similar fashion with Sec.~\ref{sec:spin2_derivation} and \cite{DoulisPhD}, one 
can use the map \eqref{eth_unit}, the expansion \eqref{spher_expansion}, and the identities \eqref{eth_on_spher_harm}
to decouple \eqref{spin2_wave_equation_GHP} on the unit sphere into seperate systems for each admissible 
pair $(l, m)$. Then, by using \eqref{GHP_derivatives} and \eqref{intrinsic_derivatives}, a coordinate 
representation of \eqref{spin2_wave_equation_GHP} can be obtained 
\begin{align}
 \begin{split}
  \label{spin2_wave_equation_final}
   & A\, C\, \partial_{tt} \Phi_\lambda + B\, (C - A)\, \partial_{tr} \Phi_\lambda - B^2\, \partial_{rr} \Phi_\lambda + B\, (2\, \sqrt{2}\, \rho -\
   2\, \sqrt{2}\, \epsilon - B')\, \partial_r \Phi_\lambda +\\ 
   & + \frac{1}{2} \left(B \left(C'-A'\right) + 2\, \sqrt{2}\, \left[(A - C)\, \rho + (2\, \lambda -5) A\, \epsilon + (2\, \lambda -3)\, C\, \epsilon \right] +\
   \dot{A}\, C + A\, \dot{C} \right) \partial_t \Phi_\lambda +\\
   & + \frac{3\, \kappa^2}{4}\, \Phi_\lambda - 2\, (\lambda ^2 - 4 \lambda - 2)\, \rho^2\, \Phi_\lambda + 8\, (\lambda -2)^2\, \epsilon^2\, \Phi_\lambda =\ 
   - \frac{\alpha_{(\lambda -1)(\lambda -2)}^2 + \alpha_{(\lambda -2) (\lambda -3)}^2}{2\, g^2\, r^2}\, \Phi_\lambda -\\
   & - \frac{\sqrt{2}\, (4 - \lambda)\, \alpha_{(\lambda -1)(\lambda -2)}\, \rho}{g\, r}\, \Phi_{\lambda +1} -\
   \frac{\sqrt{2}\, \lambda\, \alpha_{(\lambda -2) (\lambda -3)}\, \rho}{g\, r}\, \Phi_{\lambda -1},  
 \end{split}
\end{align}
where $\lambda = 0,1,2,3,4$, the functions $A, B, C, g$ are given by \eqref{metric_functions}, and the 
expansion coefficients $\rho, \epsilon$ by \eqref{spin_coef}. The above system of five equations is the 
coordinate representation of the spin-2 zero-rest-mass wave equation \eqref{spin2_wave_equation} on a 
background of the form \eqref{cyl_metric}. Notice that, as in the case of \eqref{spin2_equation_evol}, 
all the radial derivatives drop out from the above system when the equations are restricted to the cylinder 
$I$ at $r = 1$. Thus, the cylinder is again a total characteristic of \eqref{spin2_wave_equation_final}. 
  
To classify the system of second order PDEs \eqref{spin2_wave_equation_final}, we have to look at its 
symbol. The associated symmetric matrix of the principal part of the symbol of \eqref{spin2_wave_equation_final} 
reads
\[
 \left( 
  \begin{array}{cc}
   A\, C & \frac{1}{2}\, B\, (C - A) \\
   \frac{1}{2}\, B\, (C - A) & -B^2
  \end{array}
 \right)
\]
with eigenvalues 
\[
 \lambda_\pm = \frac{1}{2} \left(\omega \pm \sqrt{\omega^2 + \varepsilon^2} \right) \quad \mathrm{where} 
 \quad \omega = A\, C - B^2 \quad \mathrm{and} \quad \varepsilon = B\, (A + C).
\]
The system \eqref{spin2_wave_equation_final} is symmetric hyperbolic whenever the above eigenvalues have 
opposite sign. Obviously, $\lambda_+$ is always positive while $\lambda_-$ always negative. So, the only 
case that the system \eqref{spin2_wave_equation_final} fails to be symmetric hyperbolic is when at least 
one of the eigenvalues vanishes. For $\lambda_-$ that happens when $\varepsilon = 0$, while $\lambda_+$ 
vanishes when $\varepsilon = 0$ and $\omega = 0$. The hyperbolicity of the system \eqref{spin2_wave_equation_final} 
for the choice \eqref{f_horizontal} breaks down when $t = \pm 1$ and $r = 1$, i.e. when $\varepsilon = 0$ 
and consequently $\lambda_- = 0$. Thus, in this case, the system \eqref{spin2_wave_equation_final} is 
symmetric hyperbolic in the rectangular $\{0 < r < 1, \, |t| < 1 \}$. (Recall that the domain of symmetric 
hyperbolicity of the first order system \eqref{spin2_equation_evol} for the choice \eqref{f_horizontal} 
is exactly the same.) For the choice \eqref{f_non_horizontal}, the condition $\varepsilon = 0$ entails 
that the hyperbolicity of \eqref{spin2_wave_equation_final} breaks down ar $r = 1$. In addition, $\omega = 0$ 
leads to $t = \pm \sqrt{\pi^2 - \kappa^2}/|\kappa'|$. Combining the two last results, one concludes that 
the symmetric hyperbolicity of the system \eqref{spin2_wave_equation_final} for the choice \eqref{f_non_horizontal} 
is guaranteed in the domain $\{0 < r < 1, \, |t| < \sqrt{\pi^2 - \kappa^2}/|\kappa'| \}$. Notice that 
the domain of hyperbolicity of \eqref{spin2_wave_equation_final} is now slightly larger than the one of 
the spin-2 zero-rest-mass equation \eqref{spin2_equation_evol}. But, close to the cylinder $I$ at $r = 1$ 
their behaviour coincides as the rescaling function \eqref{kappa} tends there to zero. 

In order to prescribe initial data and constrain the evolved data on each time-slice of constant $t$, we 
supplement the five evolution equations \eqref{spin2_wave_equation_final} with the constraints \eqref{spin2_equation_constr} 
of the first order system of PDEs. The details of this procedure will be discussed in Sec.~\ref{sec:first_to_second}. 

\subsubsection{Characteristic curves}

Because of its hyperbolic nature the system \eqref{spin2_wave_equation_final} has two real characteristic 
curves. The slope of the characteristic curves for second order partial differential equations like \eqref{spin2_wave_equation_final} 
with principal part of the form $a(t,r) \partial_{tt} u(t,r) + b(t,r) \partial_{tr} u(t,r) + c(t,r) \partial_{rr} u(t,r)$ 
reads $\dd t/\dd r = (b \pm \sqrt{b^2 - 4\,a\,c})/2\,c$. Substituting $a,b,c$ according to \eqref{spin2_wave_equation_final}, 
the slope of the characteristics follows
\begin{equation*}
 \frac{\dd t}{\dd r} = - \frac{C}{B} \quad \mbox{and} \quad \frac{\dd t}{\dd r} = \frac{A}{B}. 
\end{equation*}
Interestingly, the non-linear characteristics of the spin-2 zero-rest-mass equation are identical to 
those of the spin-2 zero-rest-mass wave equation. Thus, Fig.~\ref{fig:characteristics_spin2} can be used 
to visualise them. Fig.~\ref{fig:characteristics_spin2} must be read with care though because of a small 
but substantial difference in the behaviour of the characteristics. Now, their behaviour is more universal 
in the sense that the characteristics of all the components of the spin-2 field, and not only of some 
specific components like in the case of the spin-2 zero-rest-mass equation, behave in the way depicted 
by Fig.~\ref{fig:characteristics_spin2}. Therefore, the (solid) blue and (dashed) red lines are not 
characteristics of different field components but of all components of the spin-2 field simultaneously. 
In addition, the characteristics of all five independent 
components of the spin-2 field exhibit the behaviour illustrated in Fig.~\ref{fig:characteristics_spin2}. 
One has to keep also in mind that, as mentioned above, the symmetric hyperbolic region is now a little 
bit larger, i.e., the shaded area is a little bit smaller, but close to the critical sets $I^\pm$ it 
coincides with the one represented by the green thick dashed lines of Fig.~\ref{fig:characteristics_spin2}. 
Thus, as in the case of the spin-2 zero-rest-mass equation, the entire compactified Minkowski space-time 
lies wholly into the symmetric hyperbolic region of \eqref{spin2_wave_equation_final}. The remaining 
features of Fig.~\ref{fig:characteristics_spin2} also apply, as they are, to the second order system 
of PDEs \eqref{spin2_wave_equation_final}.

\subsection{Relating the first and second order PDE systems}
\label{sec:first_to_second}

As the solution space of the second order PDE system \eqref{spin2_wave_equation} is larger than the one 
of the first order \eqref{spin2_equation}, it would be highly desirable if one could know under which 
conditions the solutions of the two systems are the same. In order to do this, we have somehow to establish 
a correspondence between the two systems that would unveil these conditions. 

Following \cite{Doulis2013,DoulisPhD}, we first define the spinor
\begin{equation}
 \label{sigma_spinor}
  \Sigma_{A'BCD} \equiv \nabla^F{}_{A'} \phi_{FBCD},
\end{equation}
which is nothing else than the l.h.s of \eqref{spin2_equation}, and then act upon it with another spinor 
covariant derivative to obtain
\[
 \nabla_{AA'} \Sigma^{A'}{}_{BCD} = \nabla_{AA'} \nabla^{FA'} \phi_{FBCD}.
\]
In accordance with the discussion in Sec.~\ref{sec:spin2_wave_derivation}, the r.h.s of the above expression 
is given by l.h.s of \eqref{spin2_wave_equation}, therefore
\begin{equation}
 \label{sigma_equation}
  \nabla_{AA'} \Sigma^{A'}{}_{BCD} = \Box \Phi_{ABCD} + \frac{3\, \kappa^2}{4}\, \Phi_{ABCD}. 
\end{equation}
Now, assuming that $\Phi_{ABCD}$ is a solution of \eqref{spin2_equation}, the spinor \eqref{sigma_spinor} 
vanishes and thus \eqref{sigma_equation} reduces to \eqref{spin2_wave_equation}, i.e. $\Phi_{ABCD}$ is 
also a solution of the spin-2 zero-rest-mass wave equation \eqref{spin2_wave_equation}.

Conversely, assuming that the spin-2 field $\Phi_{ABCD}$ is a solution of \eqref{spin2_wave_equation}, 
the expression \eqref{sigma_equation} reduces to 
\begin{equation}
 \label{sigma_equation_reduced}
  \nabla_{AA'} \Sigma^{A'}{}_{BCD} = 0. 
\end{equation}
The above system can guarantee that $\Phi_{ABCD}$ is also a solution of \eqref{spin2_equation}, i.e. 
$\Sigma_{A'BCD} = 0$, in the case that \eqref{spin2_equation} holds initially and \eqref{sigma_equation_reduced} 
is well-posed. As the condition $\Sigma_{A'BCD}|_S = 0$ can be always satisfied on an initial hyper-surface 
$S$, one has just to prove that the system \eqref{sigma_equation_reduced} is well-posed in order for $\Sigma_{A'BCD} = 0$ 
to hold throughout the evolution. To do so, we have to look at the symbol of the first order differential 
expression \eqref{sigma_equation_reduced}. Taking into account the fact that the components of \eqref{covariant_deriv} 
are $\nabla_{AA'} = (\nabla_{00'} = \mbox{\th},\nabla_{01'} = \eth,\nabla_{10'} = \eth',\nabla_{11'} = \mbox{\th}')$, 
the expansion \eqref{spher_expansion} of the components of $\Sigma_{A'BCD}$ in terms of the spin-weighted 
spherical harmonics, and the relation \eqref{intrinsic_derivatives}-\eqref{GHP_derivatives} between the 
remaining GHP-operators $\mbox{\th}, \mbox{\th}'$ and the coordinate derivatives $\partial_t, \partial_r$, 
one can write the principal part of the symbol of \eqref{sigma_equation_reduced} as follows
\[
 L^p = \sum_{\alpha = 0}^1 \mathbf{\Sigma}_\alpha(t,r) D^\alpha \quad \mathrm{with} \quad D^\alpha = (\partial_t, \partial_r), 
\]
where $\mathbf{\Sigma}_\alpha$ are $N \times N$ square matrices with $N = 8$ being the number of independent 
components of the spinor $\Sigma_{A'BCD}$. Due to the fact that the first order system of PDEs \eqref{sigma_equation_reduced} 
is symmetric under the spinor prime operation \cite{Pen&Rind1986}, each of the GHP-operators $\mbox{\th}, \mbox{\th}'$ 
will act on an equal number of components of $\Sigma_{A'BCD}$, i.e. $N/2$. Therefore, the $\mathbf{\Sigma}$-matrices 
read
\[
 \mathbf{\Sigma}_0 = \frac{1}{\sqrt{2}}\, \mathrm{diag} (\underbrace{A,\ldots,A}_{N/2},\underbrace{C,\ldots,C}_{N/2}), \qquad 
 \mathbf{\Sigma}_1 = \frac{B}{\sqrt{2}}\, \mathrm{diag} (\underbrace{1,\ldots,1}_{N/2},\underbrace{-1,\ldots,-1}_{N/2}), 
\]
where $A,B,C$ are given by \eqref{metric_functions}. As the $\mathbf{\Sigma}$-matrices are obviously 
Hermitian and $\mathbf{\Sigma}_0$ is positive definite in the range $|t| < f^{-1} (\pi/|\kappa'|)$ with 
$\dot{f} > 0$, the system \eqref{sigma_equation_reduced} is symmetric hyperbolic in this range and defines 
there a well-posed problem for $\Sigma_{A'BCD}$.
   
Summarising, we proved that the spin-2 field is a solution of both \eqref{spin2_equation} and \eqref{spin2_wave_equation} 
if and only if the initial data for both systems satisfy the constraints \eqref{spin2_equation_constr} 
of the first order system \eqref{spin2_equation}. Therefore, in the numerical implementation of \eqref{spin2_wave_equation} 
the initial data will be determined by \eqref{spin2_equation_constr} and subsequently evolved with 
\eqref{spin2_wave_equation_final}. In this way we ensure that the obtained numerical solutions are also 
solutions of \eqref{spin2_equation}.


\section{Numerical implementation and results}
\label{sec:numerics}

Here, we give a detailed description of the numerical setting that will be used to study numerically the 
systems \eqref{spin2_equation_evol}-\eqref{spin2_equation_constr} and \eqref{spin2_wave_equation_final}. 
In the present work, we mainly focus on the second order PDE system \eqref{spin2_wave_equation_final} 
as its numerical solutions have better properties than those of the corresponding first order system, 
see e.g. \cite{Kreiss&Ortiz2002}. According to \cite{Kreiss&Ortiz2002}, numerical approximations based 
on second order PDEs lead to better accuracy than the ones based on first order PDEs and, in  addition, 
prevent the occurrence of spurious high-frequency waves travelling against the characteristics. As these 
claims were also confirmed numerically in \cite{Doulis2013,DoulisPhD}, we present in Sec.~\ref{sec:num_results} 
only our findings concerning \eqref{spin2_wave_equation_final}. 

\subsection{Numerical preliminaries}
\label{sec:num_premil}

The method of lines will be used to discretize the $1+1$ system \eqref{spin2_wave_equation_final}. Accordingly, 
the PDE system \eqref{spin2_wave_equation_final} is reduced to a system of ordinary differential equations 
by discretizing the spatial coordinate $r$ with finite difference techniques. In accordance with the setting 
of Sec.~\ref{sec:blow_up} our computational domain is $D = [0,1]$. To obtain a finite representation of 
$D$ an equidistant grid $r_i = i\, h$ of grid spacing $h$ is introduced, where $i = 0,\ldots,N$, $r_N = 1$, 
and thus $h = 1/N$. The spin-2 field is discretized in a similar way $(\Phi_\lambda)_i = \Phi_\lambda(r_i)$. 
Next, we have to approximate the spatial derivatives with appropriate finite difference operators. We 
choose to use fourth order central difference operators to approximate the first and second derivatives 
appearing in \eqref{spin2_wave_equation_final} on the entire computational domain $D$ except of the grid 
points lying in the vicinity of the cylinder $I$ at $r=1$. There, we use one-sided summation by parts 
finite difference operators as in \cite{Beyer2012,Doulis2013,DoulisPhD}. (Therein, a lengthy discussion 
about the advantages of using the summation by parts operators can be found.) The reasons for this 
``inconsistency'' will become apparent in Sec.~\ref{sec:num_origin} and are related to difficulties in 
the numerical implementation of the system \eqref{spin2_wave_equation_final} at the origin $r=0$, where 
some terms of \eqref{spin2_wave_equation_final} become singular. 

Now, one has to decide how to solve the resulting semi-discrete system of ordinary differential equations. 
In order to implement the system \eqref{spin2_wave_equation_final} numerically, we reduce it to a system 
that is first order in time and second order in space by introducing the first derivatives $\Psi_\lambda = 
\partial_t \Phi_\lambda$ of the spin-2 field $\Phi_\lambda$ as additional variables. Then, the reduced system 
can be evolved in time with standard explicit fourth order Runge-Kutta schemes. When higher accuracy 
is required, especially in studies near the region $I^+$ like in the case depicted in Fig.~\ref{subfig:horizontal_charact}, 
time-step adaptive Runge-Kutta schemes that adapt the time-step to the speed of the characteristic curves 
will be employed. The adaptive time-step allows us to approach $I^+$ arbitrarily closely, but not exactly 
as the time-step then becomes arbitrarily small. 

A point that usually needs special attention is the imposition of boundary conditions at the boundaries 
of the computational domain. The mathematical framework developed in the previous sections makes the 
treatment of the boundaries a little bit easier. Specifically, as already mentioned in Sec.~\ref{sec:spin2_evolution}, 
the cylinder is a total characteristic of our system. Therefore, we are not allowed to prescribe boundary 
conditions at the points that lie there. At the boundary $r=0$, on the other side of the computational 
grid, things are a little bit more complicated. As expected the source of all our problems is related 
to the presence of $r$ in the denominator of some terms of \eqref{spin2_wave_equation_final}, which blow 
up at $r=0$. Although the spin-2 fields $\Phi_\lambda$ are expected to be regular at $r=0$, the numerical 
implementation of the (singular at the origin) equations governing their evolution is a highly non-trivial 
task. In the following section we discuss the way we chose to implement numerically the system \eqref{spin2_wave_equation_final} 
at the origin.

It is worth mentioning that the precision of Python's floats limits the spatial resolution that can be 
used in our simulations. Specifically, it imposes an upper limit to the number of grid points we can use. 
This upper limit will be determined by the initial data as follows. The spatial resolution for which Python's 
double precision, we are using, has been exceeded will mark the maximum number of grid points that can be 
used. This claim follows naturally from the fact that evolutions of initial data of higher resolution than 
the maximum allowed cannot be trusted as the numerical precision has been already exceeded on the initial 
slice. In the following, the highest resolution that we are allowed to use is $1600$ grid points in the 
non-horizontal case, Sec.~\ref{sec:num_results_non_hor}, and $600$ grid points in the horizontal case, 
Sec.~\ref{sec:num_results_hor}. 

Now, in order to check the convergence of our numerical solutions, we define the convergence rate as 
follows  
\begin{equation}
 \label{conv_rate}
  \mathrm{CR} = \frac{\log_2(E_0/E_1)}{\log_2(h_0/h_1)},
\end{equation}
where $E_0$ and $E_1$ are the normalised $l^2$ error norms for simulations of resolution $h_0$ and $h_1$, 
respectively. (Notice that $h_0 < h_1$.) The errors $E$ will be computed against the numerical simulation 
of the highest resolution, which as discussed above will consist of $1600$ grid points in the non-horizontal 
case and $600$ grid points in the horizontal case. 

The code has been written in Python and is based on the Otago relativity group's conformal field equations 
solver, which has been appropriately amended and supplemented to fit with the specific conformal problem 
we study in the present work. 

\subsection{Treatment of the origin}
\label{sec:num_origin}

At first, the $1/r$ and $1/r^2$ terms must be expressed explicitly in \eqref{spin2_wave_equation_final}. 
Observing \eqref{metric_functions} and \eqref{spin_coef} it is apparent that the quantities $g$ and $\rho$ 
are introducing the singular terms in \eqref{spin2_wave_equation_final}. As they stand, these terms are 
of the form $\csc(\frac{\pi r}{2})$ and $\csc^2(\frac{\pi r}{2})$. To express them in a more manageable 
form, we introduce the maps
\begin{equation}
 \label{r_explicit}
  \rho \mapsto S(r) - \frac{\kappa}{\sqrt{2}\, \pi\, r} \quad \mathrm{and} \quad \frac{1}{g} \mapsto r\, C(r) + \frac{\kappa}{\pi}, 
\end{equation}
where $\kappa$ is the rescaling function \eqref{kappa} and $S, C$ are the regular functions
\[
 S = \frac{r\, \kappa' + \kappa - \pi\, r\, \kappa\, \cot (\pi r)}{\sqrt{2}\, \pi\, r} \quad \mathrm{and} \quad 
 C = \kappa  \left(\csc (\pi  r)-\frac{1}{\pi  r}\right) 
\]
with $S(0) = C(0) = 0$. Expressing \eqref{spin2_wave_equation_final} through \eqref{r_explicit} in terms 
of $S$ and $C$, the $1/r$ and $1/r^2$ terms appear explicitly in \eqref{spin2_wave_equation_final}. 

In order to render regular the singular terms certain conditions must be satisfied at $r=0$. These 
regularity conditions follow naturally from the system \eqref{spin2_wave_equation_final}. There are two 
sets of regularity conditions obtained from the requirement that the coefficients of the terms $1/r$ and 
$1/r^2$ must vanish linearly and quadratically, respectively. Specifically, these conditions read
\begin{equation}
 \label{regularity_cond}
  \partial_r \Phi_\lambda = 0|_{r=0} \quad \mathrm{and} \quad 
  \left. c_1(\lambda)\, \Phi_\lambda + c_2(\lambda)\, \Phi_{\lambda + 1} + c_3(\lambda)\, \Phi_{\lambda - 1}\right|_{r=0} = 0,
\end{equation}
where $\{c_1,c_2,c_3\} \equiv \left\{(\lambda ^2 - 4 \lambda - 2 ) - (\alpha_{1,2}^2 + \alpha_{2,3}^2 )/2,\, 
(4 - \lambda)\, \alpha_{1,2},\, \lambda\, \alpha_{2,3}\right\}$ with $\alpha_{x,y} \equiv \alpha_{(\lambda -x)(\lambda -y)}$ 
and $\lambda = 0,1,2,3,4$. We must mention here that the $1/r$ terms multiplied with the functions $C,S,\kappa'$ 
are regular at the origin as these functions also vanish at $r=0$---thus, such kind of terms do not contribute 
to the first regularity condition. Notice also that with \eqref{regularity_cond} at hand one can use l'Hopital's 
rule to evaluate the singular terms. 

As it was mentioned in Sec.~\ref{sec:num_premil}, fourth order central difference operators will be used 
to approximate the first and second derivatives at (and near) $r=0$. To do so, we have to introduce a 
couple of ghost points by extending the numerical grid to negative $r$. Now, in order to evaluate 
the components of the spin-2 field at the ghost points, we take advantage of the fact that the system 
\eqref{spin2_wave_equation_final} is symmetric under a simultaneous reflection $r \mapsto -r$ and spinor 
prime operation $\Phi'_\lambda (t, -r) \mapsto \Phi_{4 - \lambda} (t, r)$. Therefore, at the grid points 
$r_0,r_1,r_2$, the first spatial derivatives will be approximated by
\begin{align}
 \label{1st_finite_diff}
  \begin{split}
   \partial_r \Phi_\lambda [r_0] &= \frac{1}{12\, h}\left(\Phi_{4- \lambda}[r_2] - 8\,\Phi_{4- \lambda}[r_1] + 8\,\Phi_\lambda[r_1] - \Phi_\lambda[r_2]\right), \\
   \partial_r \Phi_\lambda [r_1] &= \frac{1}{12\, h}\left(\Phi_{4- \lambda}[r_1] - 8\,\Phi_{4- \lambda}[r_0] + 8\,\Phi_\lambda[r_2] - \Phi_\lambda[r_3]\right), \\
   \partial_r \Phi_\lambda [r_2] &= \frac{1}{12\, h}\left(\Phi_{4- \lambda}[r_0] - 8\,\Phi_\lambda[r_1] + 8\,\Phi_\lambda[r_3] - \Phi_\lambda[r_4]\right),
  \end{split}
\end{align}
where $h$ is the grid spacing. Similarly, at the grid points $r_1,r_2$---the point $r_0$ will be discussed 
separately in the following paragraph---the second spatial derivatives will be approximated by
\begin{align}
 \label{2nd_finite_diff}
  \begin{split}
   \partial^2_r \Phi_\lambda [r_1] &= \frac{1}{12\, h^2}\left(-\Phi_{4- \lambda}[r_1] + 16\,\Phi_{4- \lambda}[r_0] - 30\,\Phi_\lambda[r_1] + 16\,\Phi_\lambda[r_2] - \Phi_\lambda[r_3]\right), \\
   \partial^2_r \Phi_\lambda [r_2] &= \frac{1}{12\, h^2}\left(-\Phi_{4- \lambda}[r_0] + 16\,\Phi_\lambda[r_1] - 30\,\Phi_\lambda[r_2] + 16\,\Phi_\lambda[r_3] - \Phi_\lambda[r_4]\right).
  \end{split}
\end{align}
 
Let's now describe the numerical implementation of the system \eqref{spin2_wave_equation_final} at $r=0$. 
Up to this point our numerical considerations are quite standard. It turns out that in order to obtain 
stable and convergent numerical solutions, the second spatial derivatives and the terms $1/r$ and $1/r^2$ 
have to be treated in a special way at the origin. 

Specifically, $\mathcal{F}/r$ terms with $\mathcal{F} \equiv \{C,S,\kappa'\}$, where $C(0) = S(0) = \kappa'(0) = 0$, 
will be replaced in accordance with l'Hopital's rule, namely $\mathcal{F}/r|_{r=0} \mapsto \partial_r \mathcal{F}|_{r=0}$. 
The remaining $1/r$ terms are proportional to $\kappa^2\,\partial_r \Phi_\lambda$ and lead to the first 
regularity condition in \eqref{regularity_cond}. According to l'Hopital's rule these terms can be evaluated 
in the following way $\partial_r \Phi_\lambda/r|_{r=0} \mapsto \partial^2_r \Phi_\lambda|_{r=0}$. Approximating 
$\partial^2_r \Phi_\lambda|_{r=0}$ with central difference operators as above, see \eqref{2nd_finite_diff}, 
did not lead to stable solutions. Interestingly, introducing the auxiliary functions $X_\lambda \equiv \partial_r \Phi_\lambda$ 
one can get stable and convergent numerical solutions by rewriting the second spatial derivatives in the 
form $\partial^2_r \Phi_\lambda|_{r=0} = \partial_r X_\lambda|_{r=0}$ and approximating them by
\begin{equation}
 \label{2nd_finite_diff_r0}
  \partial^2_r \Phi_\lambda [r_0] = \frac{1}{12\, h}\left(-X_{4- \lambda}[r_2] + 8\,X_{4- \lambda}[r_1] + 8\,X_\lambda[r_1] - X_\lambda[r_2]\right),
\end{equation}
where $X'_\lambda (t, -r) \mapsto -X_{4 - \lambda} (t, r)$ and $X_\lambda[r_i] = \partial_r \Phi_\lambda [r_i]$ 
are given by \eqref{1st_finite_diff}. Furthermore, it turns out that the stability of the solutions is 
guaranteed iff all the second spatial derivatives in \eqref{spin2_wave_equation_final}, even the ones 
not arising from a l'Hopital's rule, are approximated at $r=0$ by the above finite difference operator 
\eqref{2nd_finite_diff_r0}. 

To evaluate the $1/r^2$ terms at the origin another set of auxiliary functions $\Upsilon_\lambda \equiv \Phi_\lambda/r$ 
must be first introduced. Subsequently, each $1/r^2$ term can be written as $\Phi_\lambda/r^2 = \Upsilon_\lambda/r$ 
and the formerly quadratically singular terms now read $\left( c_1\, \Upsilon_\lambda + c_2\, \Upsilon_{\lambda + 1} + 
c_3\, \Upsilon_{\lambda - 1} \right)/r$. As before, by requiring the coefficient of $1/r$ to vanish linearly, 
the second regularity condition in \eqref{regularity_cond} can be expressed in the alternative form $\left. 
c_1\, \Upsilon_\lambda + c_2\, \Upsilon_{\lambda + 1} + c_3\, \Upsilon_{\lambda - 1} \right|_{r=0} = 0$. 
Therefore, the $1/r^2$ will be evaluated at the origin according to the rule 
\[
 \left. \frac{c_1\, \Phi_\lambda + c_2\, \Phi_{\lambda + 1} + c_3\, \Phi_{\lambda - 1}}{r^2} \right|_{r=0} \mapsto 
 \left. c_1\, \partial_r \Upsilon_\lambda + c_2\, \partial_r \Upsilon_{\lambda + 1} + c_3\, \partial_r \Upsilon_{\lambda - 1} \right|_{r=0} 
\]
with the first spatial derivative being approximated by 
\begin{equation*}
 \partial_r \Upsilon_\lambda [r_0] = \frac{1}{12\, h}\left(-\Upsilon_{4- \lambda}[r_2] + 8\,\Upsilon_{4- \lambda}[r_1] + 8\,\Upsilon_\lambda[r_1] - \Upsilon_\lambda[r_2]\right),
\end{equation*}
where $\Upsilon'_\lambda (t, -r) \mapsto -\Upsilon_{4 - \lambda} (t, r)$ and $\Upsilon_\lambda[r_i] = \Phi_\lambda [r_i]/r_i$.

\subsection{Initial data}
\label{sec:initial_data}

In accordance with the results of Sec.~\ref{sec:first_to_second}, the initial data must satisfy the conformal 
constraints \eqref{spin2_equation_constr} of the first order PDE system and be subsequently evolved with 
\eqref{spin2_wave_equation_final}. To do so, we have to bring first the constraints \eqref{spin2_equation_constr} 
in a more manageable form. Following \cite{DoulisPhD}, we construct initial data in terms of the $\Phi_2$ 
component of the spin-2 field, which can be freely specified, in a way that no differential equations have 
to be solved. 

The constraints \eqref{spin2_equation_constr} on the initial hyper-surface, where $t=0$, reduce to 
\begin{equation}
 \label{constraints_t=0}
  2\, \kappa\, \partial_r \Phi_{\lambda+1} - \alpha_{1,2}\, \pi\, C\, \Phi_\lambda - \frac{\alpha_{1,2}\, \kappa}{r}\, \Phi_\lambda - 6\, \sqrt{2}\, \pi\, S\, \Phi_{\lambda+1} +\
  \frac{6\, \kappa}{r}\, \Phi_{\lambda+1} - \alpha_{0,1}\, \pi\, C\, \Phi_{\lambda+2} - \frac{\alpha_{0,1}\, \kappa}{r}\, \Phi_{\lambda+2} = 0, 
\end{equation}
where $\alpha_{x,y} \equiv \alpha_{(\lambda -x)(\lambda -y)}$ and $\lambda=0,1,2$. Clearly, the system 
\eqref{constraints_t=0} is underdetermined and symmetric under the oparation $\Phi_\lambda(0,r) \mapsto \Phi_{4-\lambda}(0,r)$. 
One way of solving \eqref{constraints_t=0} is by specifying freely  two of the components of the spin-2 
field and expressing the remaining three in terms of them. Alternatively, inspired by the aforementioned 
symmetry of \eqref{constraints_t=0}, one can require that $\Phi_\lambda(0,r) = \Phi_{4-\lambda}(0,r)$ 
holds on the initial hyper-surface. In this way, the unknowns have been reduced to three and the independent 
equations constraining them to two. Therefore, by specifying freely the, e.g., $\Phi_2$ component of the 
spin-2 field, a specific family of solutions of \eqref{constraints_t=0} can be obtained algebraically:
\begin{align}
 \label{initial_data}
  \begin{split}
   & \Phi_1(0,r) = \Phi_3(0,r) = \frac{\kappa\, x\, \partial_r \Phi_2 + 3\, \pi \left(\kappa - \sqrt{2}\, S\, x \right) \Phi_2}{\alpha_0\, \pi\, (\kappa + x\, C)}, \\
   & \Phi_0(0,r) = \Phi_4(0,r) = \frac{1}{\pi^2\, \alpha_0\, \alpha_2 (\kappa + x\, C)^3} \left\{ 2\, \kappa^2\, x^2\, (\kappa + x\, C)\, \partial^2_r \Phi_2 + \right.\\
   & + 2\, \kappa\, x \left(x^2 \left(C \left(\kappa' - 6\, \sqrt{2}\, \pi\,  S \right) - \kappa\, C' \right) + 6\, \pi\, \kappa\, x \left(C - \sqrt{2}\, S \right) +\
   7\, \pi\, \kappa^2 \right) \partial_r \Phi_2 -\\
   & - 6\, \pi \left[\kappa\, x^2 \left(C' \left(\kappa -\sqrt{2}\, S\, x \right) + \sqrt{2}\, S'\, (\kappa + x\, C)- \left(C + \sqrt{2}\,S \right) \kappa' \right) - \right.\\
   & \left.\left.- \pi \left(6\, C\, S^2\, x^3 + 6\, \kappa\, S\, x^2 \left(S - \sqrt{2}\, C \right) + \kappa^2\, x \left(2\, C - 7\, \sqrt{2}\, S \right) + 3\, \kappa^3 \right) +\
   \frac{\pi}{6}\, \alpha_0^2\, (\kappa + x\, C)^3\right] \Phi_2 \right\},
  \end{split}
\end{align}
where $x \equiv \pi\, r$. Hence, having specified the field $\Phi_2$ explicitly, the rest components of 
the spin-2 field can be computed algebraically from the system \eqref{initial_data}. 

In the following, we will choose the field $\Phi_2$ to be initially a bump function of the form  
\begin{equation}
 \label{initial_data_phi_2}
  \Phi_2(0,r) = \left\{
   \begin{array}{cccc} 
    \left(4\, \frac{(r - a)(r - b)}{(b - a)^2} \right)^{16} & , & a \leq r \leq b & \\
                                0                           & , & \,\,\, r > b \,\,\,\, \mathrm{and} & r < a 
   \end{array} \right.
\end{equation}
centered at $r = (a + b)/2$. Then, the rest of the components of the spin-2 field are also bump functions 
and can be computed from \eqref{initial_data}. In this work, we choose $a = 0$ and $b = 1$, which guarantees 
that the domain where \eqref{initial_data_phi_2} is non-trivial coincides with our computational domain, 
i.e. $0 \leq r \leq 1$, and that the initial data vanish at the boundary. 

In addition, because the system \eqref{spin2_wave_equation_final} is second order in time, the values of 
the first temporal derivatives $\Psi_\lambda$ of the spin-2 field's components must be also specified on 
the initial hyper-surface. The evolution equations \eqref{spin2_equation_evol} of the first order system 
of PDEs, evaluated at $t=0$, will be used for this purpose. The values of the fields on the r.h.s of 
\eqref{spin2_equation_evol} can be evaluated from \eqref{initial_data} and \eqref{initial_data_phi_2}. 

\subsection{Results}
\label{sec:num_results}

The initial data constructed in the previous section will be evolved now with the system of second order 
PDEs \eqref{spin2_wave_equation_final} in the two distinct conformal compactifications of Minkowski space-time 
presented in Fig.~\ref{fig:compactification_cyl}. We report that our findings concerning the advantages 
of using the second order system \eqref{spin2_wave_equation_final} instead of the first order system of PDEs \eqref{spin2_equation_evol} 
to evolve the above initial data, confirm the respective ones in \cite{Doulis2013,DoulisPhD,Beyer2014b}. 
Namely, evolutions with \eqref{spin2_wave_equation_final} lead to better accuracy and suppress the appearance 
of the high-frequency waves that travel against the characteristics and spoil the convergence of our numerical 
simulations.

\subsubsection{Non-horizontal representation}
\label{sec:num_results_non_hor}

\begin{figure}[htb]
 \centering
  \subfigure[]{
   \includegraphics[scale = 0.44]{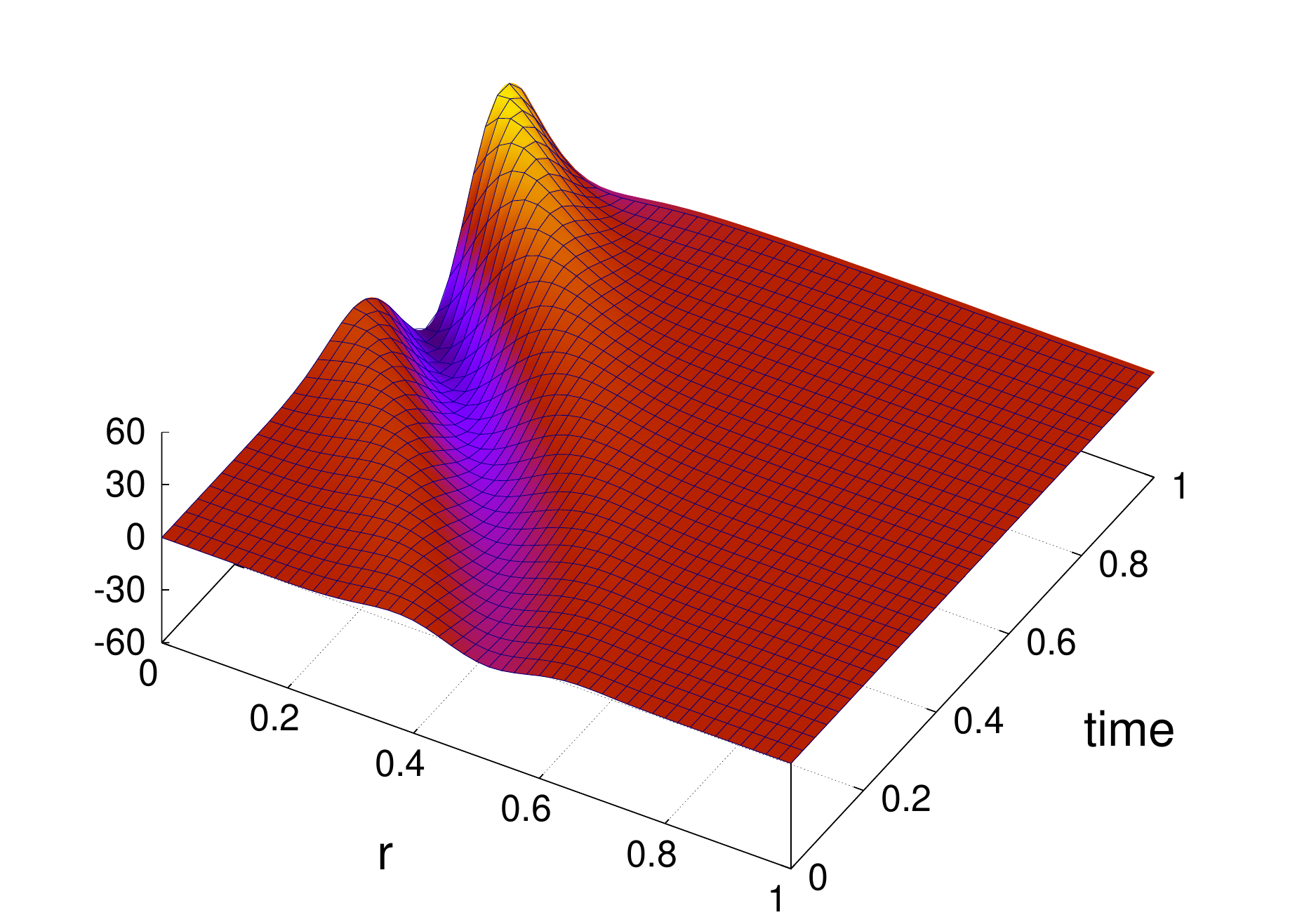}
   \label{subfig:num_sol_non_horiz_phi0}
  }
  \hspace{-1.6cm}
  \subfigure[]{
   \includegraphics[scale = 0.44]{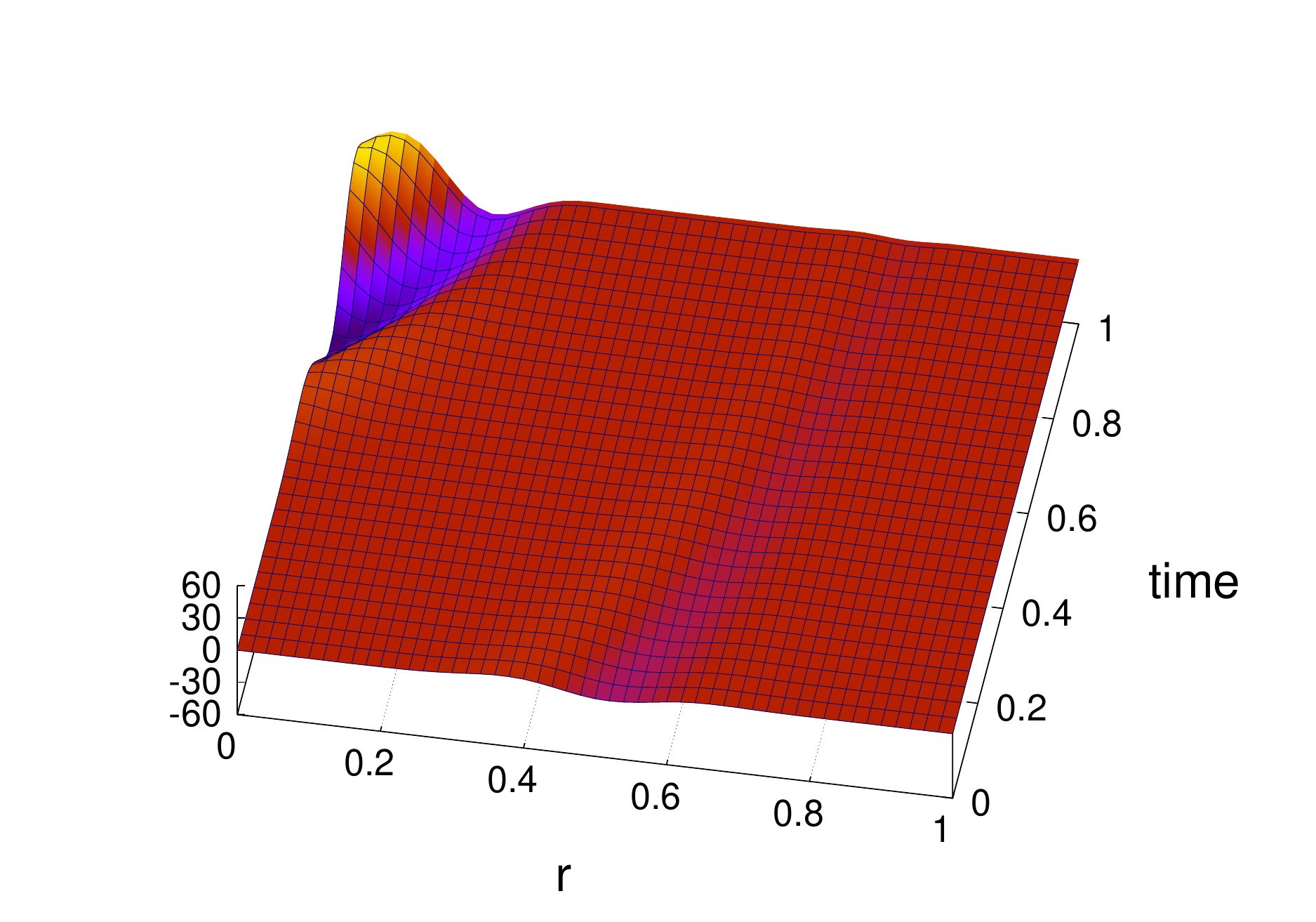}
   \label{subfig:num_sol_non_horiz_phi4}
  }
 \caption{The numerical solutions for (a) $\Phi_0$ and (b) $\Phi_4$ resulting from the evolution of the 
 initial data \eqref{initial_data}-\eqref{initial_data_phi_2} in the representation of Fig.~\ref{subfig:non_horizontal_compact}. 
 Notice that along the timeline of the origin $\Phi_0 (t, 0) = \Phi_4 (t, 0)$ always  holds. This is just 
 a mere consequence of the symmetry of the system \eqref{spin2_wave_equation_final} under transformations 
 of the form $\Phi_\lambda (t, -r) \mapsto \Phi_{4 - \lambda} (t, r)$, which were used in Sec.~\ref{sec:num_origin} 
 to regularise the singular terms of \eqref{spin2_wave_equation_final}.}
 \label{fig:num_sol_non_horiz}
\end{figure}

First, we present our results for the non-horizontal case of Fig.~\ref{subfig:non_horizontal_compact}, 
where $f = 2\, t$. The lowest non-trivial mode $l=2$ will be considered here. Thus, \eqref{initial_data} 
and \eqref{initial_data_phi_2} together with \eqref{spin2_equation_evol}, evaluated at $t=0$, for the 
choice $l=2$ will be our initial data. We evolve these data with \eqref{spin2_wave_equation_final} using 
an explicit fourth order Runge-Kutta scheme with a constant time-step $\Delta t = \mathcal{C}\, h$, where 
$\mathcal{C}$ is the so-called CFL number. The results of the present section are obtained with $\mathcal{C} = 0.05$. 
Recall also that the boundary at $r=1$ does not require boundary conditions as it is a total characteristic 
of \eqref{spin2_wave_equation_final}, while the boundary at $r=0$ is treated in the way described in detail 
in Sec.~\ref{sec:num_origin}. The resulting numerical solutions for the components $\Phi_0$ and $\Phi_4$ 
of the spin-2 field are presented in Fig.~\ref{fig:num_sol_non_horiz}. Clearly, $\Phi_0$ moves towards 
the origin while $\Phi_4$ moves in the opposite direction towards the cylinder $I$ at $r=1$. The other 
components behave in a similar way.

Although, the numerical solutions displayed in Fig.~\ref{fig:num_sol_non_horiz} look quite smooth and 
stable, we have to conduct further tests to conclude with certainty that they are stable and convergent. 

\begin{figure}[htb]
 \centering
  \subfigure[]{
   \includegraphics[scale = 0.58]{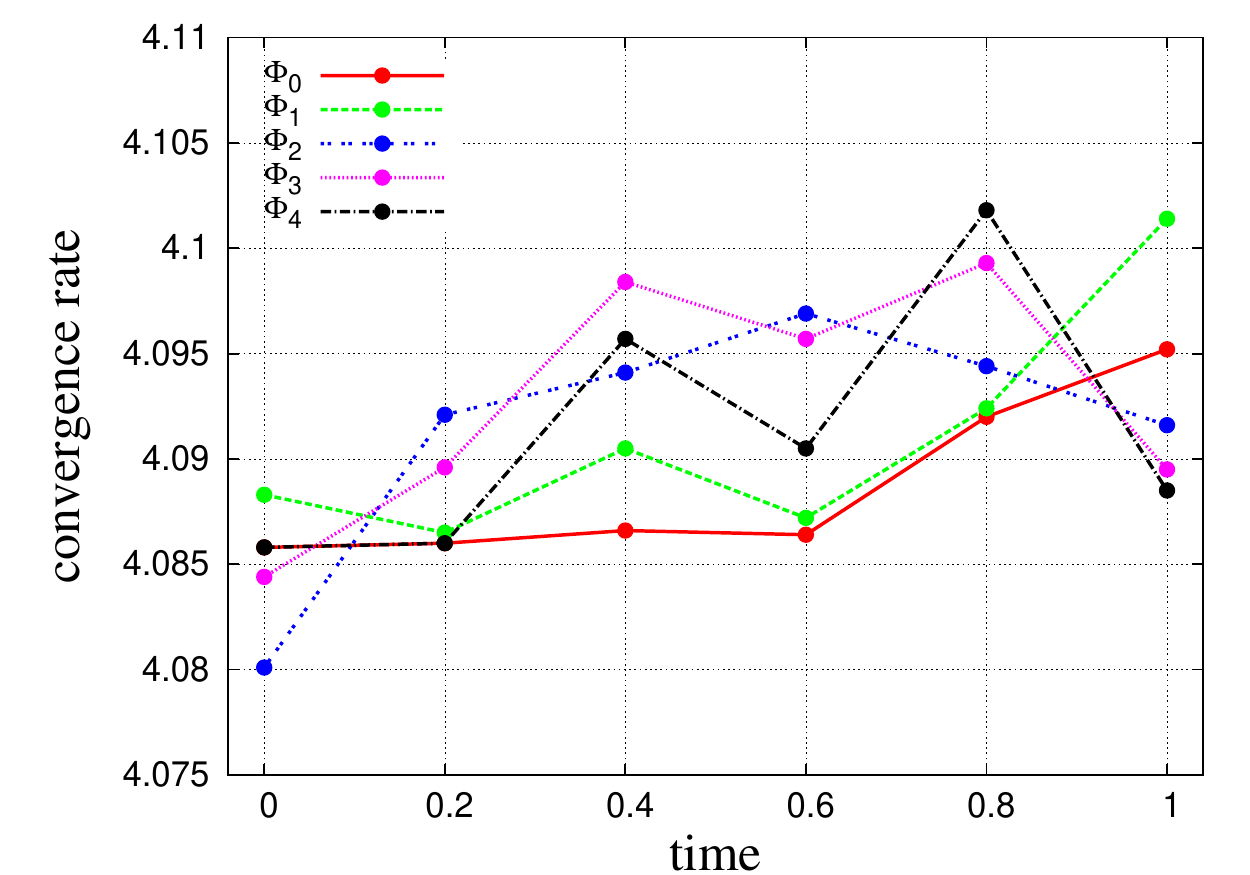}
   \label{subfig:conv_rates_non_hor}
  }
  \hspace{0.3cm}
  \subfigure[]{
   \includegraphics[scale = 0.58]{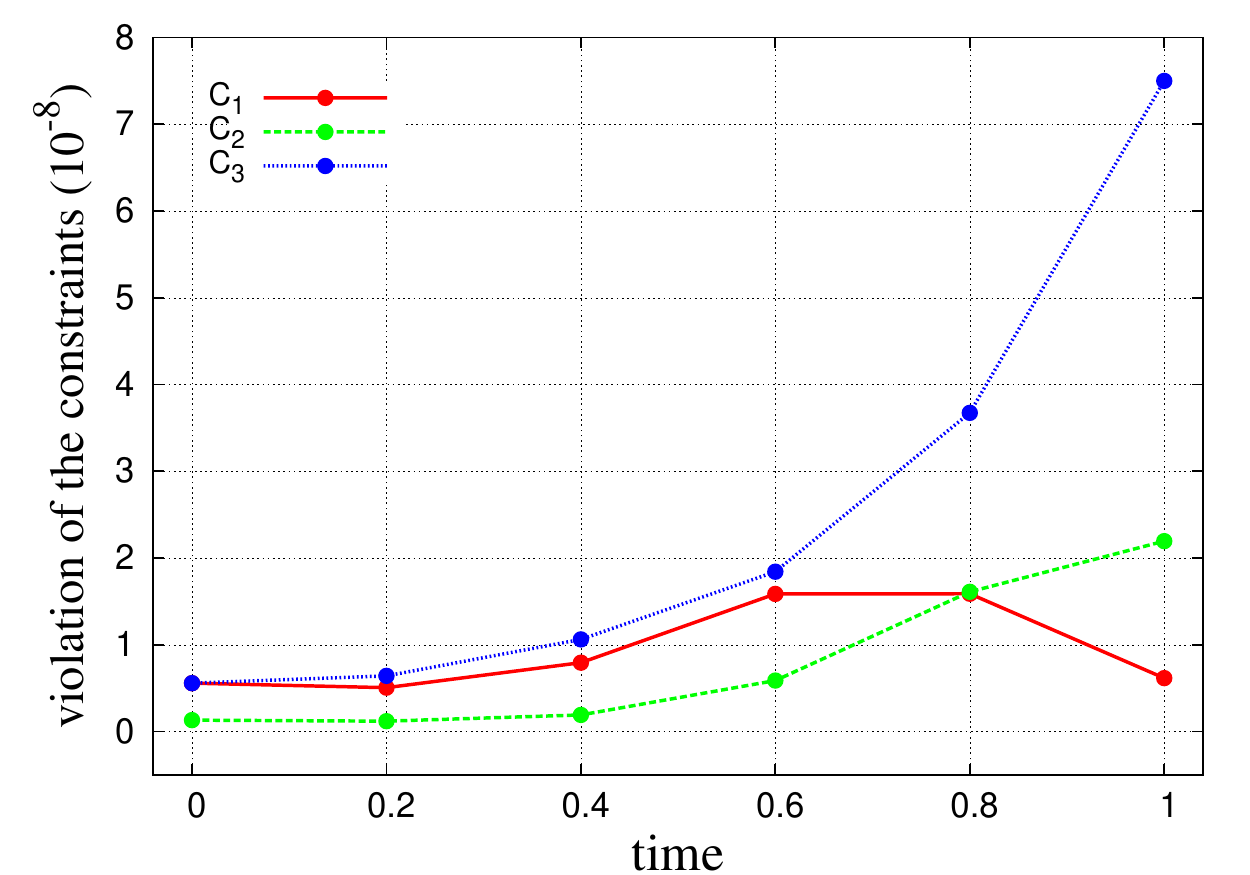}
   \label{subfig:viol_constr_non_hor}
  }
 \caption{For the numerical solutions of Fig.~\ref{fig:num_sol_non_horiz}, we present the temporal evolution 
 (a) of the convergence rates o each one of the components of the spin-2 field and (b) of the violation 
 of the vanishing of the constraint quantities \eqref{spin2_equation_constr}. The critical set $I^+$ at 
 $t=1$ can be successfully reached, but as expected we cannot go beyond it, i.e. we always maintain $t<1$.}
 \label{fig:test_non_horiz}
\end{figure}

At first, we can look at their convergence rates \eqref{conv_rate}. The behaviour of the convergence 
rates with time for each component of the spin-2 field is illustrated in Fig.~\ref{subfig:conv_rates_non_hor}. 
It is clearly visible that during the evolution the convergence rates of all the components are a little 
bit above $4$ --- a result that is in good agreement with the expected fourth order convergence of our numerical 
scheme. This expectation follows naturally from the fact that the time integration is performed with a 
fourth order Runge-Kutta method and the spatial derivatives are approximated with fourth order finite 
difference operators.
 
Another way to test our numerical solutions is by checking if the vanishing of the constraint quantities 
\eqref{spin2_equation_constr} is satisfied during the evolution. As it was mentioned previously, it is 
almost impossible not to violate the quantities \eqref{spin2_equation_constr} while prescribing numerically 
initial data on the initial hyper-surface. We cannot avoid introducing initially some error into our system. 
Fortunately, as it was proved at the end of Sec.~\ref{sec:spin2_evolution}, the structure of the subsidiary 
system of \eqref{spin2_equation_constr} does not allow these errors to grow exponentially during the 
evolution. Thus, we have to check if the numerical solutions of Fig.~\ref{fig:num_sol_non_horiz} confirm 
this expectation. To do so, we evaluate at each time-step the three constraint quantities and compute 
their normalised $l^2$ norms along the whole computational domain. Fig.~\ref{subfig:viol_constr_non_hor} 
depicts the behaviour of the three constraint quantities during the evolution. Taking into account that 
the initial violation of the constraints is, depending on the constraint quantity, between $10^{-9}$ and 
$10^{-8}$, then the data of Fig.~\ref{subfig:viol_constr_non_hor} clearly indicate that the violation is 
contained at these levels during the evolution. This extremely pleasant feature is a consequence of the 
linear nature of \eqref{spin2_wave_equation_final} and of the comparatively short period of evolution. 
In addition, the convergence of the constraint quantities with increasing resolution has been also checked. 
Our findings agree with the expected 4th order convergence to zero. 

It is expected that logarithmic singularities will develop when our numerical simulations reach the critical 
set $I^+$ at $t=1$. (Recall that beyond $I^+$ the domain of non-hyperbolicity of \eqref{spin2_wave_equation_final} 
starts, see Fig.~\ref{subfig:non_horizontal_charact}.) Our findings confirm this expectation. Namely, 
because we use an explicit Runge-Kutta scheme, $I^+$ can be successfully reached, but going beyond leads 
immediately to instabilities and code crash. This means that the part of Minkowski space-time beyond $t=1$ 
cannot be covered by our numerical simulations. So, it is not possible to cover the whole of Minkowski 
space-time in the representation of Fig.~\ref{subfig:non_horizontal_compact}. But, there is a possibility 
to achieve this in the horizontal representation of Fig.~\ref{subfig:horizontal_compact} discussed in the 
following section. 

\subsubsection{Horizontal representation}
\label{sec:num_results_hor}

Let us now turn to the horizontal representation of Fig.~\ref{subfig:horizontal_compact}, where the whole 
of Minkowski space-time is represented as a conformally equivalent region of the Einstein static universe 
with a rectangular shape in the chosen coordinates. This representation is numerically quite advantageous 
as the critical set $I^+$, future null $\scri^+$ and time-like $i^+$ infinity are located at the same time-slice 
$t=1$. Therefore, with the last time-step, that takes us to $t=1$, we not only reach $I^+$ but also $\scri^+$ 
and $i^+$. Thus, we do not have to go beyond $t=1$ to cover the whole Minkowski space-time. But this comes 
at a price, the speed of the characteristic curves at $t=1$, see Fig.~\ref{subfig:horizontal_charact}, 
becomes infinite---which makes our endeavour to reach $t=1$ extremely challenging. In the following, we 
investigate numerically the possibility of reaching $t=1$ in this setting. 

\begin{figure}[htb]
 \centering
  \subfigure[]{
   \includegraphics[scale = 0.44]{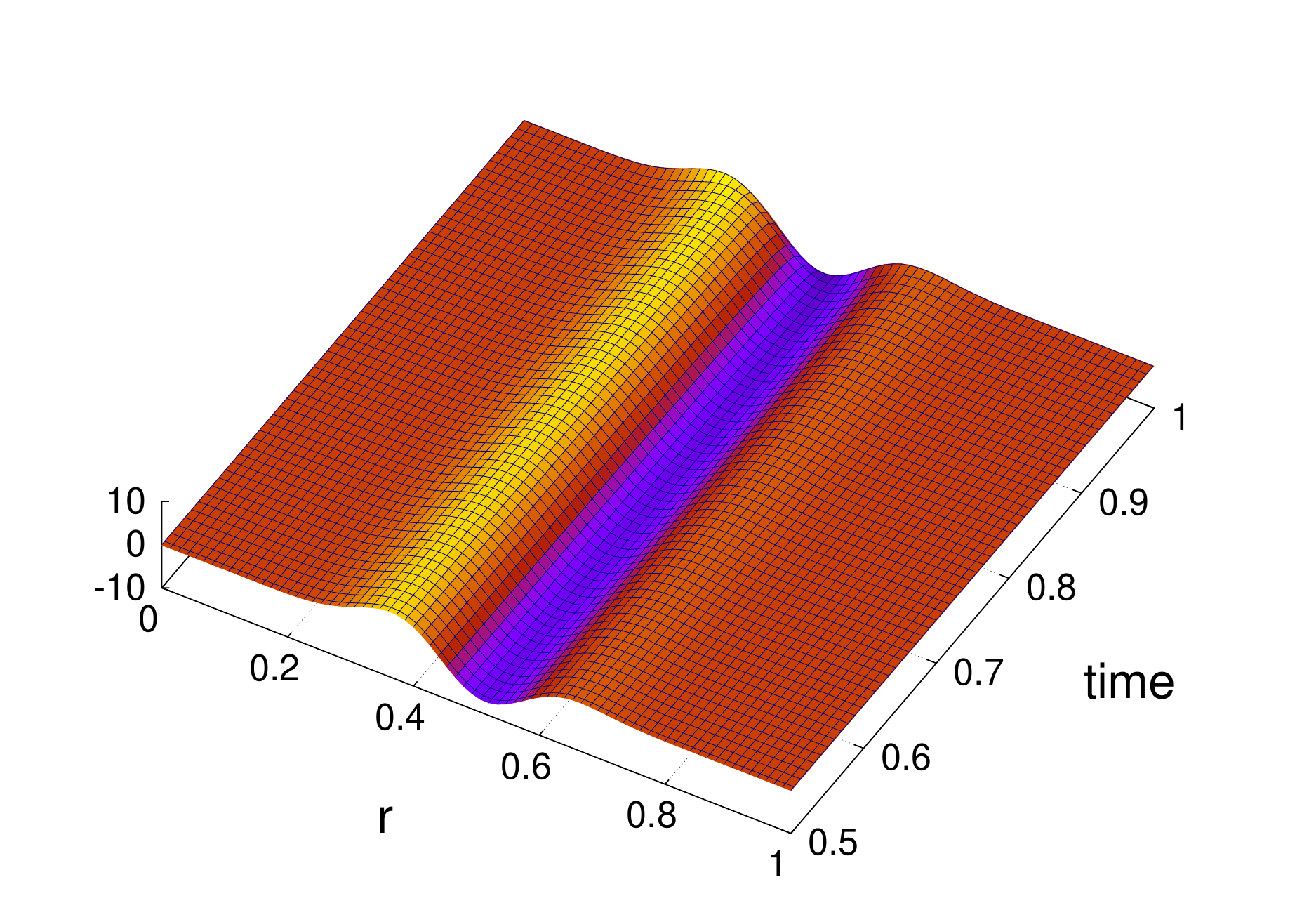}
   \label{subfig:num_soln_horiz_phi0}
  }
  \hspace{-1.6cm}
  \subfigure[]{
   \includegraphics[scale = 0.44]{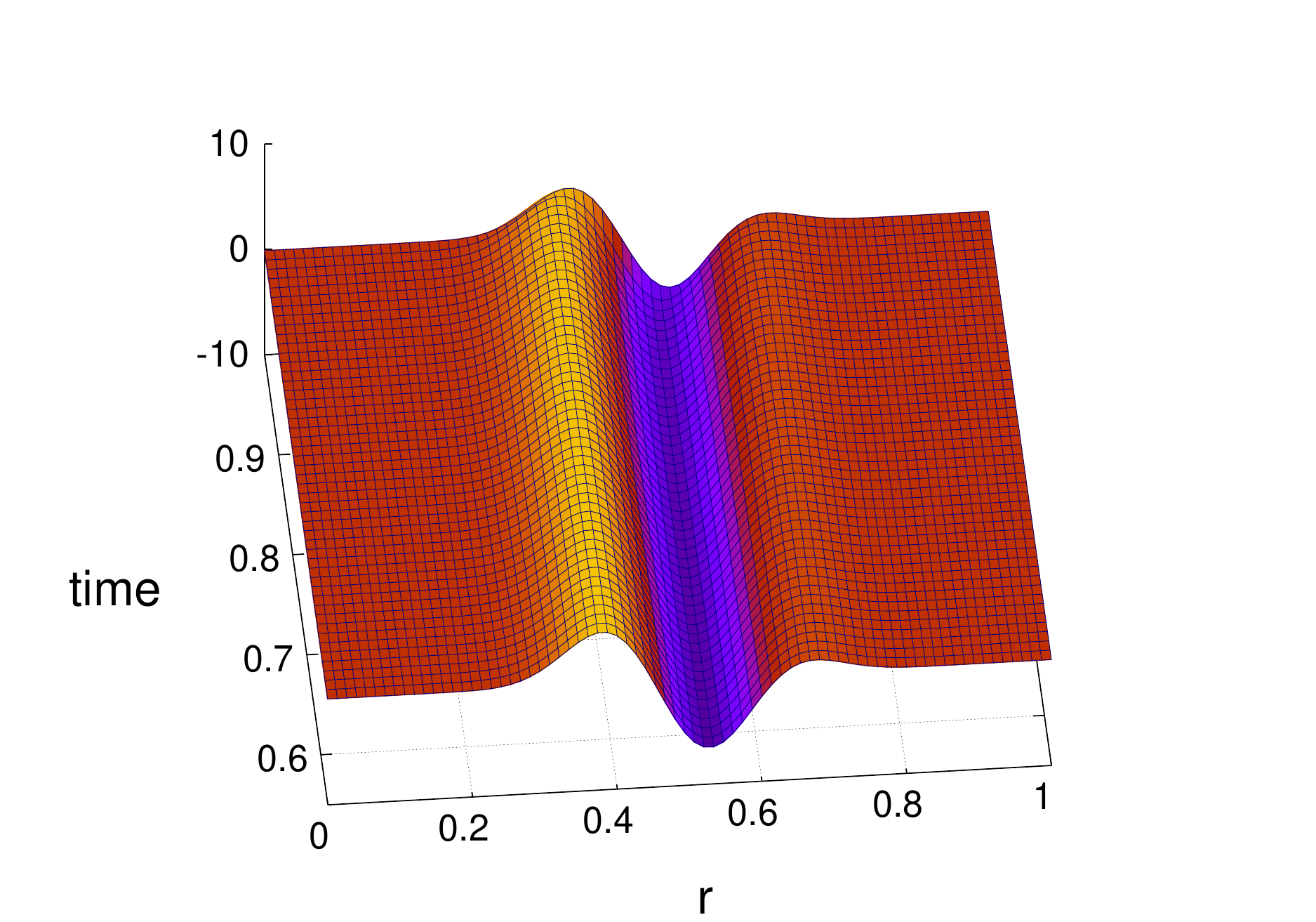}
   \label{subfig:num_sol_horiz_phi4}
  }
 \caption{The numerical solutions for (a) $\Phi_0$ and (b) $\Phi_4$ resulting from the evolution of the 
 initial data \eqref{initial_data}-\eqref{initial_data_phi_2} in the representation of Fig.~\ref{subfig:horizontal_compact}. 
 Notice the increase of the propagation speed at late times $t \approx 1$ attributed to the abrupt shift 
 of the orientation of the characteristic curves there, see Fig.~\ref{subfig:horizontal_charact}.}
 \label{fig:num_sol_horiz}
\end{figure}

As above, our initial data consist of \eqref{initial_data} and \eqref{initial_data_phi_2} together with 
\eqref{spin2_equation_evol}, evaluated at $t=0$, for the lowest non-trivial mode $l=2$ and the choice 
\eqref{f_horizontal}. Again, we use an explicit fourth order Runge-Kutta scheme of constant time-step 
with CFL number $\mathcal{C} = 0.1$. The resulting numerical solutions for the components $\Phi_0$ and 
$\Phi_4$ of the spin-2 field are presented in Fig.~\ref{fig:num_sol_horiz}. Notice that while $\Phi_0$ 
and $\Phi_4$ move slowly towards the origin and the cylinder, respectively, at late times this indolent 
movement is accelerated. A look at the characteristic curves of Fig.~\ref{subfig:horizontal_charact} 
suffices to explain this behaviour. The rapid shift of the orientation of the characteristics from almost 
vertical to almost parallel at late times is responsible for the observed increase of the propagation 
speed of the evolved data. The remaining components behave in a similar way. 

By inspection of Fig.~\ref{fig:num_sol_horiz}, the obtained solutions are evidently smooth and stable 
during the whole evolution. At first sight, this seemingly doubtful result which implies that our numerical solutions 
are smooth and stable even at $t=1$, where the characteristic speed becomes infinity, can be attributed 
to the explicit Runge-Kutta scheme we are using and to the extremely steep characteristics of the horizontal 
representation, see Fig.~\ref{subfig:horizontal_charact}. Because of the former, the evolution equations 
are actually never evaluated at $t=1$. The latter now delays the violation of the CFL condition, resulting 
from the increase of the characteristic speed, to very late times. When appropriately combined, these 
two features can lead to the smooth and stable solutions of Fig.~\ref{fig:num_sol_horiz}. It is noteworthy 
that when the characteristic curves are less steep than the ones we are using, then $t=1$ cannot be reached 
in a stable way and the solutions blow up there. So, the steeper the characteristic curves are, the better 
our numerical results. The numerical factor in \eqref{f_horizontal} controls the steepness of the characteristics; 
this explains why the specific value appearing in \eqref{f_horizontal} was chosen.

\begin{figure}[htb]
 \centering
  \subfigure[]{
   \includegraphics[scale = 0.6]{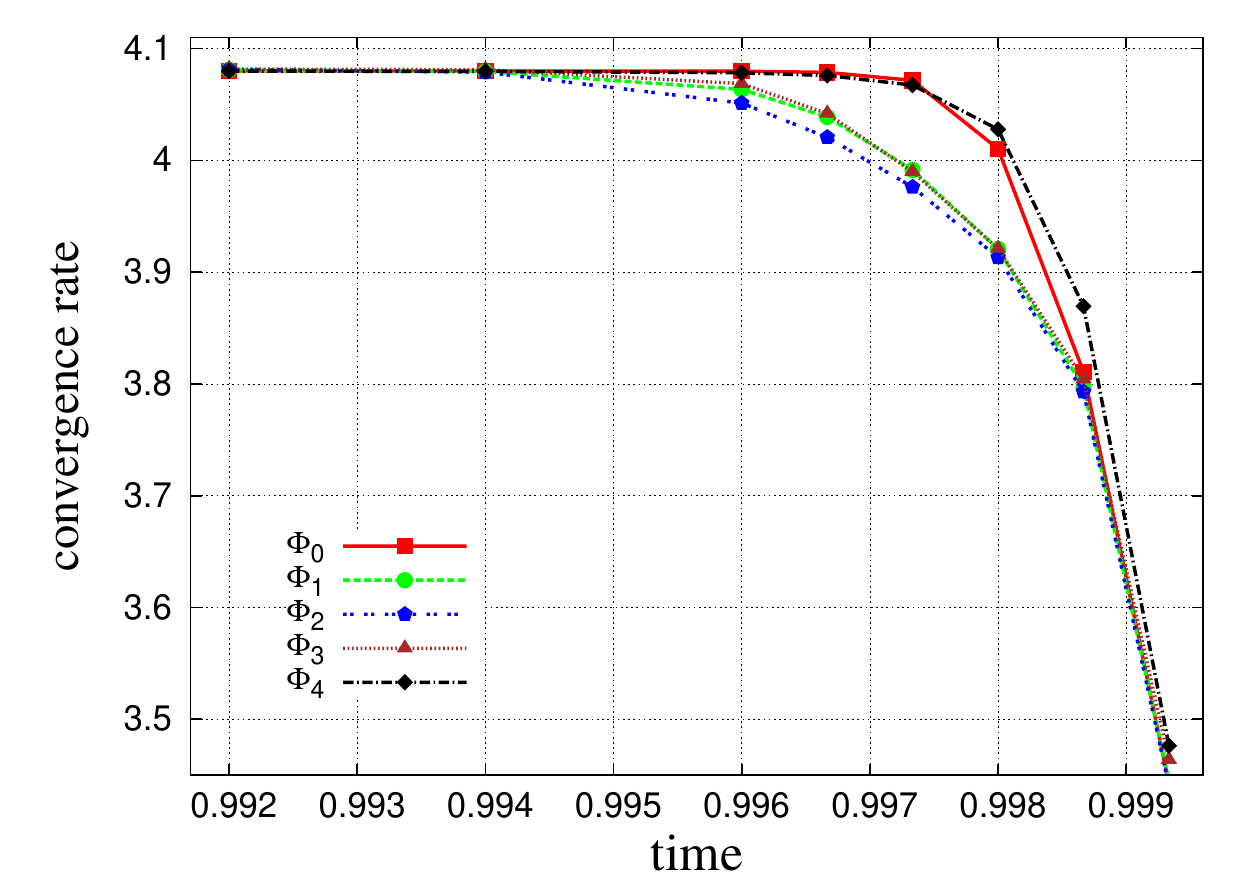}
   \label{subfig:conv_rates_hor}
  }
  \hspace{-0.2cm}
  \subfigure[]{
   \includegraphics[scale = 0.6]{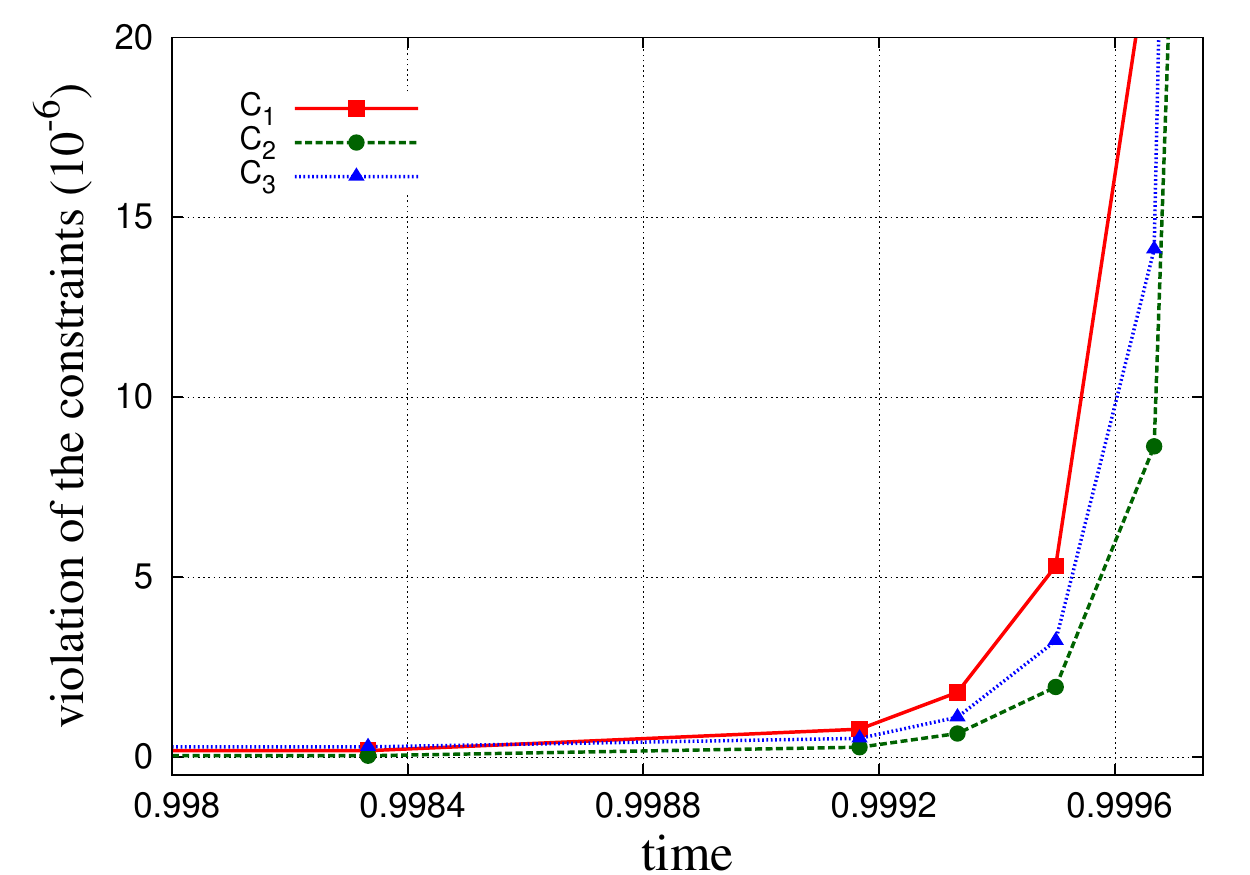}
   \label{subfig:viol_constr_hor}
  }
 \caption{For the numerical solutions of Fig.~\ref{fig:num_sol_horiz}, the time dependence (a) of the 
 convergence rates o each one of the components of the spin-2 field and (b) of the violation of the 
 vanishing of the constraint quantities \eqref{spin2_equation_constr} is presented. The observed loss 
 of convergence and non-preservation of the constraints is caused  by the rapid increase of the characteristic 
 speed at late times, which in turn leads to the violation of the CFL condition.}
 \label{fig:test_horiz}
\end{figure}

To study the convergence properties of the solutions of Fig.~\ref{fig:num_sol_horiz}, we take a look at 
their convergence rates \eqref{conv_rate}. Fig.~\ref{subfig:conv_rates_hor} illustrates the convergence 
rates for each component of the spin-2 field as a function of time. We have zoomed in to the interesting 
region where $t \rightarrow 1$ as for earlier times $t < 0.992$ the convergence rates are well above $4$. 
For late times we lose convergence while approaching $t=1$ and end up with convergence rates close to unity 
at $t=1$. The observed loss of convergence can be ascribed to the violation of the CFL condition 
caused by the increase of the characteristic speed at late times, see Fig.~\ref{subfig:horizontal_charact}. 
Thus, by decreasing the CFL number $\mathcal{C}$, the loss of convergence can be significantly reduced 
and postponed to even later times, but cannot be avoided altogether as we approach $t=1$ where the characteristic 
speed becomes infinite. 

Now, let's study the behaviour of the vanishing of the constraint quantities \eqref{spin2_equation_constr} 
during the evolution. Fig.~\ref{subfig:viol_constr_hor} depicts the temporal evolutions of the constraint 
quantities $C_\lambda$ with time. The initial violation of the constraints is of the order of $10^{-7}$ 
and is maintained at this level, as can be seen in Fig.~\ref{subfig:viol_constr_hor}, until the quite 
late time $t \approx 0.999$. Thereafter, the constraints are increasingly violated. This is again a 
consequence of the violation of the CFL condition that follows from the increase of the characteristic 
speed while approaching $t=1$. Decrease of the CFL number $\mathcal{C}$ delays the phase of increasing 
violation of the constraints but cannot suppress it. 

\begin{figure}[htb]
 \centering
  \subfigure[]{
   \includegraphics[scale = 0.6]{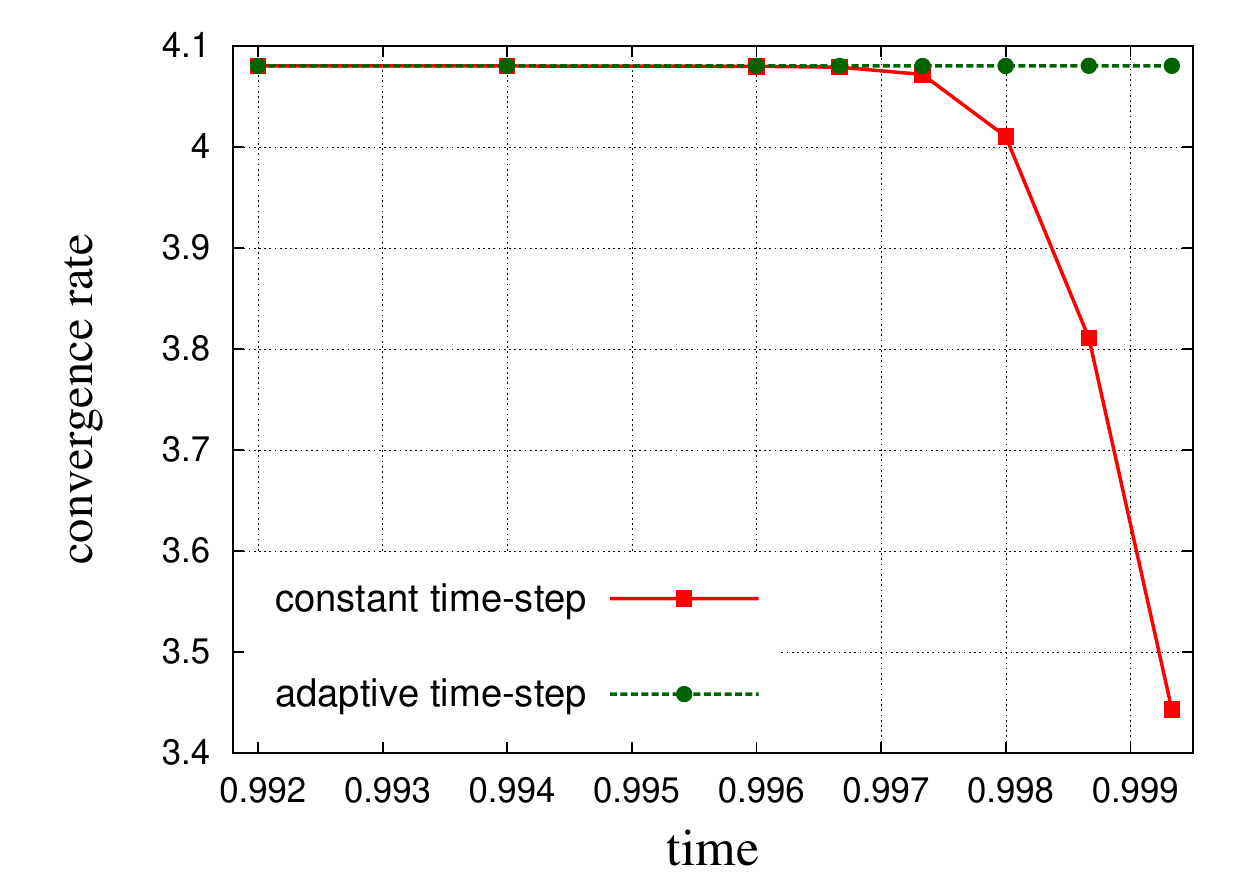}
   \label{subfig:conv_rates_hor_adapt}
  }
  \hspace{-0.2cm}
  \subfigure[]{
   \includegraphics[scale = 0.6]{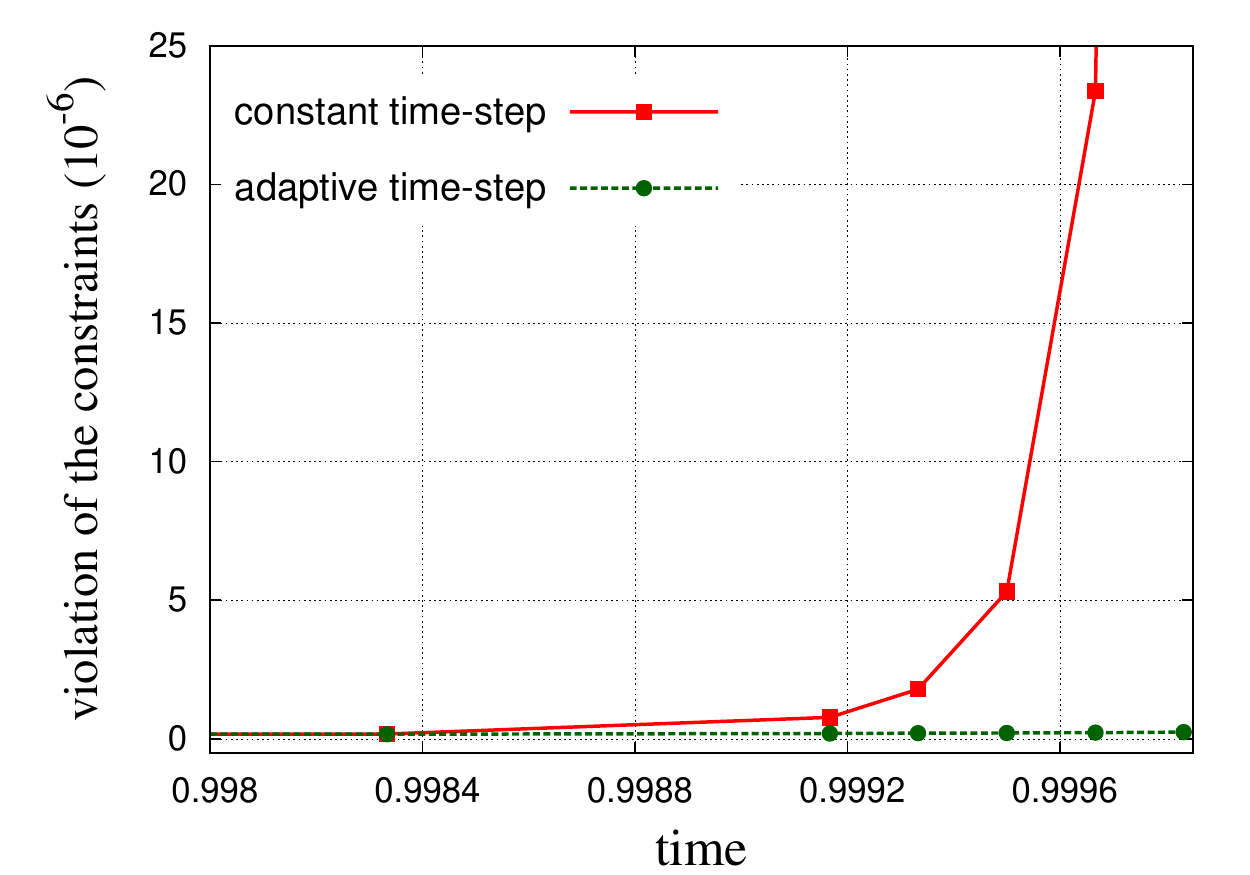}
   \label{subfig:viol_constr_hor_adapt}
  }
 \caption{For the numerical solutions of Fig.~\ref{fig:num_sol_horiz}, we present the behaviour with 
 time (a) of the convergence rate of the worst converging component $\Phi_0$ of the spin-2 field and (b) 
 of the violation of the vanishing of the mostly violated constraint quantity $C_1$. In each figure, we 
 compare the results of Fig.~\ref{fig:test_horiz} obtained with a constant time-step to the ones obtained 
 with an adaptive time-step. Obviously, the use of an adaptive time-step restores the respective quantities 
 to their expected values and enables us to reach $t=1$ to a distance of merely $10^{-12}$.}
 \label{fig:test_horiz_adapt}
\end{figure}

Recall that all our results in the present section have been obtained using a time-integrator that marches 
at a constant pace throughout the evolution. So, when using a constant time-step, our findings above show 
that, although our numerical solutions do not blow up at $t=1$, $t=1$ cannot be reached without a loss 
of convergence and a considerable violation of the constraints that are caused by the unavoidable increase 
of the characteristic speed while $t \rightarrow 1$. If the latter is really the reason for the underperformance 
of our code close to $t=1$, then by using an explicit Runge-Kutta scheme with an adaptive time-step, $t=1$ 
could be approached arbitrarily close without losing convergence and increasingly violating the constraints. 
A time-integrator of this type adjusts the time-step, and consequently the CFL number $\mathcal{C}$, 
according to the magnitude of the characteristic speed in a way that the CFL condition is always satisfied. 
In Fig.~\ref{fig:test_horiz_adapt} the convergence rates and the behaviour of the constraint quantities 
\eqref{spin2_equation_constr} resulting from an evolution with an adaptive time-step are compared to the 
corresponding results of Fig.~\ref{fig:test_horiz} obtained with a constant time-step. Clearly, the use 
of an adaptive time-step restores the convergence rates and the preservation of the constraints to their 
expected values. In addition, for the simulation of highest resolution, here $600$ grid points, we managed 
to advance to $t \approx 0.999999999999$, just $10^{-12}$ from $t=1$. We could not get closer to $t=1$ 
as the time-step becomes of the order of $10^{-17}$ exceeding Python's double precision. In principle, 
using higher quadruple or octuple precision can get us even closer to $t=1$. 


\section{Discussion}
\label{sec:discussion}

In this work it was shown that the generalised conformal field equations \cite{Friedrich1998} can be used 
to study gravitational perturbations on the whole of Minkowski space-time $\mathbb{M}$ and not only in the 
regions close to space-like infinity $i^0$. 

This has been achieved by slightly diverging from Friedrich's original formulation where space-like infinity 
$i^0$ is first placed at the origin by a coordinate inversion and then blown up to a cylinder by an appropriate 
rescaling of the resulting space-time. Here, a different strategy was followed. Instead of inverting the 
coordinates, we first conformally compactified $\mathbb{M}$ into the Einstein static universe $\mathbb{E}$, 
see Sec.~\ref{sec:compactification}, and then by appropriately rescaling the resulting conformal metric 
\eqref{mink_comp_metric}, space-like infinity was blown up to a cylinder in the spirit of Friedrich, 
see~\eqref{cyl_metric}. The free functions $\kappa(r)$ and $f(t)$, introduced by the rescaling, control the 
shape and the location of $i^0$ and $\scri$. Here, we chose to work in the two representations of 
Fig.~\ref{fig:compactification_cyl}. The former representation is an example of the general setting where 
the temporal positions of $i^+,i^0$ and $\scri^+$ on $\mathbb{M}$ are distinct, while the latter is quite 
special as all of $i^+,i^0,\scri^+$ are positioned on the same time-slice. 

The price to pay for including the whole of $\mathbb{M}$ into the computational domain is that some terms 
of the generalised conformal field equations, in both their representation as a system of first \eqref{spin2_equation_evol} 
and second \eqref{spin2_wave_equation_final} order PDEs, are singular at $r=0$. Note, however, that this 
is not due to a deficiency of the conformal rescaling but due to the use of polar coordinates adapted to 
the spherical symmetry of the background Minkowski space-time. Therefore, although the spin-2 field is 
regular at the origin, the numerical implementation of the equations governing its dynamics is highly involved. 
In Sec.~\ref{sec:num_origin}, the numerical implementation of the system \eqref{spin2_wave_equation_final} 
at the origin is described in detail. 

In the non-horizontal representation of Fig.~\ref{subfig:non_horizontal_compact} it was possible to reach 
the critical set $I^+$ at $t=1$ without loss of convergence and with the constraint quantities \eqref{spin2_equation_constr} 
preserved, see Sec.~\ref{sec:num_results_non_hor}. This was made possible because of the finite speed of 
the characteristic curves at $I^+$, see Fig.~\ref{subfig:non_horizontal_charact}, and of the explicit 
Runge-Kutta scheme we are using. But any attempt to go beyond $I^+$, i.e. to enter the domain of non-hyperbolicity 
of \eqref{spin2_wave_equation_final}, leads as expected almost immediately to code crash. Therefore, it 
is not possible to cover parts of $\mathbb{M}$ lying beyond the time-slice $t=1$ in this representation. 

In the horizontal representation of Fig.~\ref{subfig:horizontal_compact} the whole of Minkowski space-time 
has been restricted conformally between the time-slices $t=-1$ and $t=1$ (which are, in fact, null hyper-surfaces) 
that go through the critical sets 
$I^-$ and $I^+$, respectively. This feature leaves open the possibility of performing a global simulation 
of $\mathbb{M}$, namely evolving data from past $i^-$ all the way to future time-like infinity $i^+$. 
This possibility was extensively investigated in Sec.~\ref{sec:num_results_hor}, the main source of our 
difficulties here is related to the fact that the speed of the characteristics becomes infinite at $t=1$, 
see Fig.~\ref{subfig:horizontal_charact}. A constant time-step throughout the evolution can get us to 
$t=1$ in a smooth way, see Fig.~\ref{fig:num_sol_horiz}, but with a considerable loss of convergence and 
violation of the constraints, see Fig.~\ref{fig:test_horiz}. We can get around this by using an adaptive 
time-step. In this case how close we can get to $t=1$ depends on the available computational precision. 
With Python's double precision we managed to approach $t=1$ to a distance of only $10^{-12}$ with the 
expected 4th order convergence and preservation of the constraints, see Fig.~\ref{fig:test_horiz_adapt}. 

Having shown that our formulation can be successfully applied for the Minkowski space-time, we can move 
on and apply it to space-times subject to less restrictive symmetry conditions.


\section{Acknowledgments}
JF would like to thank the Department of Mathematics at the University of Oslo for hospitality. Part of
this research was supported by the European Research Council through the FP7-IDEAS-ERC Starting Grant
scheme, project 278011 STUCCOFIELDS.


\bibliography{origin_Mink}

\end{document}